\documentclass{pasj00}
\draft
\usepackage{longtable}
\usepackage{pdflscape}
\usepackage{multirow}
\usepackage[figuresright]{rotating}
\usepackage{multicol}
\usepackage{multirow}
\usepackage{booktabs}
\usepackage{threeparttable}

\begin{document}
\SetRunningHead{Author(s) in page-head}{Running Head}

\title{Spectroscopic properties of the dwarf nova-type cataclysmic variables observed by LAMOST}



%
 \author{Han \textsc{Zhongtao},\altaffilmark{1,2,3,4,5}
         Boonrucksar \textsc{Soonthornthum},\altaffilmark{2}
         Qian \textsc{Shengbang},\altaffilmark{1,4,5,6}
         Fang \textsc{Xiaohui},\altaffilmark{1,4,6}
         Wang \textsc{Qishan},\altaffilmark{1,4,6}
         Zhu \textsc{Liying},\altaffilmark{1,4,5,6}
         Dong \textsc{Aijun}\altaffilmark{3}
        and Zhi \textsc{Qijun}\altaffilmark{3}
         }
\altaffiltext{1}{Yunnan Observatories, Chinese Academy of Sciences (CAS), P. O. Box 110, 650216 Kunming, China}
 \email{zhongtaohan@ynao.ac.cn}
\altaffiltext{2}{National Astronomical Research Institute of Thailand, Higher Education Science Research and Innovation, 260 Moo 4, T. Donkaew, A. Maerim, Chiangmai, 50180, Thailand}
\altaffiltext{3}{Guizhou Provincial Key Laboratory of Radio Data Processing, School of Physics and Electronic Sciences, Guizhou Education University, Guiyang 550001, China}
\altaffiltext{4}{Key Laboratory of the Structure and Evolution of Celestial Objects, Chinese Academy of Sciences, P. O. Box 110, 650216 Kunming, China}
\altaffiltext{5}{Center for Astronomical Mega-Science, Chinese Academy of Sciences, 20A Datun Road, Chaoyang District, Beijing, 100012, P. R. China}
\altaffiltext{6}{University of Chinese Academy of Sciences, Yuquan Road 19\#, Sijingshang Block, 100049 Beijing, China}


\maketitle

\begin{abstract}
Spectra of 76 known dwarf novae from the LAMOST survey were presented. Most of the objects were observed in quiescence, and about 16 systems have typical outburst spectra. 36 of these systems were observed by SDSS, and most of their spectra are similar to the SDSS spectra. 2 objects, V367 Peg and V537 Peg, are the first to observe their spectra. The spectrum of V367 Peg shows a contribution from a M-type donor and its spectral type could be estimated as M3-5 by combining its orbital period. The signature of white dwarf spectrum can be seen clearly in four low-accretion-rate WZ Sge stars. Other special spectral features worthy of further observations are also noted and discussed. We present a LAMOST spectral atlas of outbursting dwarf novae. 6 objects have the first outburst spectra, and the others were also compared with the published outburst spectra. We argue that these data will be useful for further investigation of the accretion disc properties. The HeII $\lambda$4686 emission line can be found in the outburst spectra of seven dwarf novae. These objects are excellent candidates for probing the spiral asymmetries of accretion disc.
\end{abstract}s

\section{Introduction}

Dwarf novae are the most populous subtype of the cataclysmic variables (CVs) and show frequent and dramatic outbursts of typically 2$-$8 mag (Warner 1995).
In general, the outbursts last for about a few days to weeks. The recurrence time varies widely between several days to tens of years for different systems, correlated with the duration of the outbursts. The dwarf nova outburst is reasonably well understood as a release of gravitational energy, caused by a thermal instability of accretion discs (Smak 1971; Osaki 1974; H\={o}shi 1979; Osaki 1996; Lasota 2001). More specifically, the disc-instability model suggested that during quiescence the matter is accumulated within the accretion disc, and then rapidly dumped onto the white dwarf during outburst via a thermal instability. Approximately 50\% of the known CVs belong to dwarf nova-type systems, and many short-period CVs ($P_{orb}\leq2$ hr) are the SU UMa-type dwarf novae, which show less frequent but brighter superoutbursts. The superoutburst phenomenon of SU UMa stars can be explained by the combination of thermal and tidal instabilities of the accretion disc, called thermal-tidal instability model (Osaki 1989). However, the details regarding these theories are quite complicated and still poorly understood.

Early surveys discovered dwarf novae are based on their characteristic colors or emission lines such as the Palomar-Green Survey (Green et al. 1986; Ringwald 1993) and the Hamburg Quasar Survey (Hagen et al. 1995; G\"{a}nsicke et al. 2002; Aungwerojwit 2005). Due to the outbursts, the dwarf novae are generally identified as optical transients. This makes the transient surveys very useful in the dwarf nova discovery. A large number of dwarf novae have been discovered or identified from these surveys, e.g., the Catalina Real-time Transient Survey (CRTS; Drake et al. 2009), the Mobile Astronomical System of the Telescope Robots (MASTER; Lipunov et al. 2010) and the All Sky Automated Survey (ASAS; Pojmanski 1997). However, most historical dwarf novae were nearby blue, bright systems or fainter objects experiencing a larger amplitude outburst owing to observational bias inherent from such surveys. Obviously, these samples suffer from strong observational selection effects. Therefore, it is necessary to obtain a large homogeneous sample of dwarf novae in order to understand their evolution and outburst. More recently, a more comprehensive detection method from the Sloan Digital Sky Survey (SDSS) was used to find a more uniform and a truer representation sample of CVs. To date, more than 100 dwarf novae have been identified by using the spectroscopic database of SDSS and follow-up observations (see Szkody et al. 2011, 2014 and references therein). Compared with the photometric data during outburst, however, their outburst spectra is still relatively lacking.
The photometric observations can provide a large amount of information pertaining to the fundamental outburst properties such as outburst strength, duration, recurrence time and the statistical correlations between them. However, for more details on the intrinsic properties of the system's components and the outburst process, the spectroscopic observations are even more useful.

The optical spectra of the dwarf novae are the superposition of the emission from each component. Typical characteristics in quiescence are the presence of prominent hydrogen Balmer and neutral helium emission lines. It is thought that these emission lines are resulted in the accretion disc. Meanwhile, there are other lines from a few ionized helium and heavier elements. Some systems with double-peaked emission feature usually have relatively high orbital inclination and are important candidates of eclipsing CVs. Moreover, the ratio of H$\beta$ to HeII $\lambda$4686 is very large in the quiescence spectra. However, the spectral features in outburst are changed from the emission lines to absorption lines (Joy 1960; Warner 1976). Some intensive studies of the spectra change from the quiescent to outburst state have been given in the literatures (Warner 1995 and references therein for more details; Neustroev et al. 2017). In particular, the HeII $\lambda$4686 emission during outburst is regarded as the best tracer of spiral structure in the accretion disc (Harlaftis et al. 1999). The spiral arms have been detected in several systems around peak of outburst by using the trailed spectrogram of HeII, such as IP Peg (Harlaftis et al. 1999; Baptista et al. 2005), EX Dra (Joergens et al. 2000), SS Cyg (Steeghs 2001), U Gem (Groot 2001) and WZ Sge (Kuulkers et al. 2002). In addition, a comprehensive study by Morales-Rueda \& Marsh (2002) found a number of candidates for exploring the spiral structure. We also expect to find the similar outburst spectra in this work.

The Large Sky Area Multi-Object Fiber Spectroscopic Telescope (LAMOST) is a quasi-meridian reflecting Schmidt telescope (Cui et al. 2012). Its field of view is 5 degrees and the effective aperture is about 4 meters. This telescope was designed to conduct a wide-field spectral survey using 16 250-channel spectrographs.
Each spectrograph has a blue (3700-5900{\AA}) and a red (5700-9000{\AA}) channel CCD camera with a low resolving power of $\sim$ 1800. The telescope is capable of getting 4000 spectra simultaneously in one exposure. As of now, the LAMOST survey has released the full fifth data (DR5), a part of sixth data (DR6) and a part of seventh data (DR7). The total spectra number has been more than 10 million. These data are extremely valuable for studying the variable stars and close binaries (e.g., Qian et al. 2017, 2018a, 2018b, 2019a, 2019b). About 76 dwarf novae were observed by LAMOST from 2011 October 24 to 2018 Dec 31 and a total of 131 spectra were obtained. Of $\sim$30\% of these objects have outburst spectra, which provide clues into the changes of each components during outburst.
Here, we present the LAMOST spectra of 76 dwarf novae and use these data to analyze their spectroscopic properties at outburst and quiescence state.

\section{Data}

The present work is based on the LAMOST DR5, the first five months of DR6 and the first three mouths of DR7. The spectroscopic data were reduced by using the two-dimensional (2D) pipeline software of LAMOST (Luo et al. 2015) which corrected, extracted and calibrated the spectra, and combined two wavelength regions (i.e. red and blue). Note that the relative flux calibration was performed for the LAMOST spectra (Song et al. 2012). More than 70\% spectra presented here have the signal-to-noise ratio (SNR) $\geq$ 10. The reduced data were stored in LAMOST database, and some specific users were allowed to download them.

Over the years, the number of CVs is increasing rapidly as the carried out of the sky surveys and programs. All of these objects have been gathered into a catalogue by Ritter \& Kolb (2003, 2014). By 31 Dec 2015, the catalogue has been updated to the final edition (RKcat 7.24\footnote{https://wwwmpa.mpa-garching.mpg.de/RKcat/index.html}) which contains 1429 CVs. 576 of these CVs were classified as dwarf novae. In addition, 105 new dwarf novae were discovered by SDSS (e.g. Szkody et al. 2011, 2014 and references therein). Here we cross-matched all known dwarf novae with the over 10 million LAMOST spectra to date, and find that about 10\% dwarf novae (76) were observed by LAMOST. The observation log for dwarf novae is given in Table 1, which lists LAMOST name, orbital period, subtype, plan ID, target observation date, the Modified Julian Day (MJD), the Seeing, the exposure time, the Local Modified Julian Minute at the start time of each plate (MJM) and their activity state. Of the 76 dwarf novae in Table 1, 30 (40\%) are observed more than once. To assess the activity state of these spectra, we checked the AAVSO (American Association of Variable Star Observers\footnote{https://www.aavso.org/}) light curves for each object.
Figure 1 shows the spectra of U Gem in both quiescence and outburst states. The time that such spectra were taken are marked by the arrows of different colors (see the inset of the lower panel). We can see clearly the change from quiescence state with emission lines to outburst peak with absorption lines. The quiescent spectrum in the upper panel shows prominent hydrogen Balmer and helium emission lines as well as CaII emission in the infra-red, thought to be due to its accretion disc. In addition, the TiO absorption bands from M-type donor are also seen in this spectrum.
The outburst spectra in the lower panel display clear absorption features, typical characteristic of a system in outburst or a high state of accretion.
Note that, at outburst maximum of U Gem, the flux has raised by a factor of 8 in the red and 40 in the blue. Moreover, the strong HeII $\lambda$4686 emission line could be seen in the outburst spectra. This is commonly attributed to the irradiation from a thick accretion disc (Smak 2001; Ogilvie 2002), or is caused by the spiral structures (Morales-Rueda \& Marsh 2002).

\section{Analysis and Results}
Of the 76 dwarf novae observed by LAMOST, about 20 have the typical spectra that are in or near outburst by examining the AAVSO data, and other systems were observed in quiescence. Figs. 2-6 display the typical quiescent spectra and Fig. 7 shows several special quiescent spectra, which were obtained before or after outburst. The outburst spectra were displayed in Figs. 8-11, where we also showed the corresponding AAVSO light curves. Noted that only 2 of our objects (i.e. V367 Peg and V537 Peg) have never obtained the spectra, and about 36 ($\sim47.4$\%) have been observed by SDSS. Even though the the overlap of both the SDSS and LAMOST is large, there are still 2 systems at the different states. Therefore, LAMOST spectra remain very valuable and useful in CVs, and the two surveys are complementary. Table 2 lists the measurements for the equivalent widths (EWs) and the full width at halfmaximum (FWHM) of several prominent spectral lines. In this section, our main purpose is to study the spectral properties of these dwarf novae and to discuss some particular objects and lines.

\subsection{Dwarf novae in quiescence}

Based on the characteristics of spectral lines and the fluxes in the red and blue, the spectra in quiescence were classified. The observations and their activity states were summarized in Table 1.
During quiescence, the dwarf nova has an optically thin disc with a large temperature gradient from the inner region to outer region. This produces a relatively flat continuum plus some emission lines in the optical spectrum (Hellier 2001). Typical disc emission lines include the Balmer series (H$\alpha$, H$\beta$, H$\gamma$ etc.), the neutral helium (HeI) series, the calcium triplet lines (CaII $\lambda$8498, 8542, 8662) and a few heavier element lines (e.g., FeII $\lambda$4924, 5169, 5317). Most of the quiescent spectra presented in Figs. 2-7 were dominated by these emission lines. We note that both the EW and FWHM values in quiescence can change in the same object (see Table 2).

However, apart from the accretion disc, there are a late-type (K or M) donor star and a white dwarf. Thus, a dwarf nova spectrum also should in principle contain the spectral features from these components, which characterise the blue continuum superposed broad absorption lines and strong atomic or molecular absorption bands (e.g., Mg near $\lambda$5200, TiO near $\lambda$7100). The spectral feature of the M star donor can be seen in several objects which have been reported to contain the M type stars previously (except for V367 Peg). These objects are GY Cnc (G{\"a}nsicke et al. 2000), IP Peg (Martin et al. 1987), HH Cnc (Thorstensen et al. 2015), SDSS J100658.41+233724.4 (Southworth et al. 2009), SDSS J105550.08+095620.4 (Thorstensen \& Skinner 2012), U Gem (Stauffer et al.1979), V405 Peg (Thorstensen et al.2009), V1239 Her(Littlefair et al.2006) and VZ Sex (Mennickent et al.2002).  Also, there is a weak absorption feature around $\lambda$8180 that can be found in some of such objects. This feature is the sodium infrared doublet line (Na I $\lambda$8183/8195{\AA}), which has been widely identified previously in K and M dwarfs. In general, the weak absorption lines at longer wavelengths are derived from the molecules of cool donor atmosphere. Furthermore, two well-known CVs, GK Per and SS Cyg, show typical characteristic of the K-type star. Their donor has also been confirmed to be K star for a long time (see Morales-Rueda et al. 2002 and Bitner et al. 2007 for summaries).  However, the white dwarf spectrum is only visible in the systems with low accretion rates and faint accretion discs (G\"{a}nsicke et al. 2009). There are four objects in our sample: EG Cnc, EZ Lyn, PQ And and V355 UMa, in which the broad absorption wings around Balmer lines are clearly visible. This feature has been seen in the past published spectra of these objects (e.g. Patterson et al. 1998; Szkody et al. 2006; Schwarz et al.2004; G{\"a}nsicke et al. 2006). They have the orbital periods of 84.97 min, 84.63 min, 80.64 min and 82.53 min respectively, typical of WZ Sge-type stars (Kato et al 2004, 2015; Patterson et al. 2005).

Besides the normal spectral lines, the HeI $\lambda$5876 emission line together with the sodium D doublet absorption lines (Na D $\lambda$5890/5896{\AA}) are particularly notable. 10 of our targets show this feature, they are: EM Cyg, EI Psc, GK Per, HH Cnc, IU Leo, NZ Boo, SDSS J124417.89+300401.0, SDSS J153634.42+332851.9, SS Cyg and VZ Sex. The Na D absorption lines are a common feature in the CV spectra and they most likely originate from a combination of interstellar absorption and/or the donor star. 
    However, there are two bad regions near $\lambda$6000{\AA} and $\lambda$8500{\AA} in the LAMOST spectra due to the data-reduction issues (Han et al. 2018). All unusual features around these two wavelengths are not credible. Nevertheless, the different objects still have their unique features. We also find that 2 of our sample were observed through LAMOST and SDSS in different states. Here we provide the notes and comments on the particular systems below.

\emph{BF Eri}. The spectrum taken on 2015 Dec 5 shows an unusual HeI $\lambda$5876 emission line profile. Compared with the normal profile, this emission line has an extended structure on right side, consisting of multi-peaks. This peculiar shape was not found in previous spectroscopic studies (Schachter et al. 1996; Neustroev \& Zharikov 2008). However, the other line profiles do not show any similar features, so this profile is not to be trusted.

\emph{SDSS J080846.19+313106.0}. The LAMOST spectrum of this object was observed on 2012 Jan 26 and shows strong hydrogen emission lines and flat continuum, typical of a CV in quiescence. However, its SDSS spectrum exhibited the feature of broad absorption lines along with weak emissions in the core, which is typical of a dwarf nova during outburst (Szkody et al. 2004).

\emph{SDSS J105550.08+095620.4}. The LAMOST spectrum in quiescence was obtained on 2017 Apr 27 and displays clear feature of a M-type donor. Although an $M3\pm2$ secondary star was presented, the M-dwarf   is less prominent in the published quiescence spectra (Thorstensen \& Skinner2012; Thorstensen et al. 2016). The SDSS spectrum showed strong blue continuum and weak emission lines, indicating an outbursting dwarf nova (Szkody et al. 2011).

\begin{landscape}
\begin{table*}
\tiny
\caption{LAMOST observation log for dwarf novae. The orbital periods and sub-types were taken from the Ritter-Kolb catalogue (RKcat 7.24)and the International Variable Star Index database (VSX).}
 \begin{center}
 \begin{tabular}{lllllllllll}\hline\hline
Targets                       &LAMOST name                       &$P_{orb}$(d)   &Type    &Planid               &Date             &MJD              &Seeing($''$)        &Exp.T(s)       &MJM       &States        \\\hline
GK Per                        &LAMOST J033112.00+435415.2        &1.996803               &DN/IP        &VB053N42V2           &2012/12/25          &2456286            &3.6           &600           &81053152           &quiescence              \\
V537 Peg                      &LAMOST J224340.70+305520.0        &0.42234                &DN              &HD224534N302633B01   &2014/11/11          &2456972            &3.2           &1500          &82040807          &decline?                \\
SY Cnc                        &LAMOST J090103.34+175356.2        &0.382375               &ZC           &K2QSOB01             &2017/12/18          &2458105            &5.5           &1500          &83672750         &decline                    \\
&        &               &           &K2QSOB02             &2017/12/21          &2458108            &2.7           &1500          &83677071         &decline                    \\
&        &               &           &K2QSOB01             &2018/01/08          &2458126            &5.6           &1500          &83702996         &near peak                    \\
&        &               &           &K2QSOB01             &2018/01/10          &2458128            &4.0           &1500          &83705846         &early decline                    \\
&        &               &           &K2QSOB01             &2018/01/11          &2458129            &5.1           &1500          &83707288         &early decline                    \\
&        &               &           &K2QSOB01             &2018/01/12          &2458130            &2.1           &1500          &83708718         &decline                    \\
&        &               &           &KII090344N200347V01  &2018/12/03          &2458455            &4.3           &600          &84176975          &outburst peak                    \\
IU Leo                        &LAMOST J105756.33+092314.9        &0.376308               &UG           &HD110713N101244B     &2015/01/01          &2457023            &6.5           &500           &81118120         &decline                   \\
&        &               &           &HD104915N103242B02   &2017/02/05          &2457789            &3.6           &1500          &83217630         &quiescence                   \\
HS 0218+3229                  &LAMOST J022133.47+324323.8        &0.297229661                &UG        &HD022353N335128B01   &2018/01/18          &2458136            &3.0           &1500          &83716965           &$-$               \\
EM Cyg                        &LAMOST J193840.11+303028.3        &0.290909               &ZC        &HD193711N282959V01   &2016/05/26          &2457534            &2.7           &600           &82850592           &late decline              \\
&        &               &        &HD193711N282959V01   &2016/09/09          &2457640            &2.0           &600           &83002791           &near outburst              \\
SS Cyg                        &LAMOST J214242.80+433509.8        &0.275130               &UG           &HD213239N421743V01   &2014/10/18          &2456948            &3.0           &600           &82006376         &quiescence                    \\
BF Eri                        &LAMOST J043929.95-043558.8        &0.270880               &UG        &EG043159S042253M01   &2015/12/05          &2457361            &3.0           &1800          &82601210           &quiescence              \\
V344 Ori                      &LAMOST J061518.94+153059.4        &0.234               &UG           &GAC093N17M1          &2014/10/26          &2456956	           &3.2           &1800          &82018263        &quiescence                    \\
&        &               &           &GAC092N13M1          &2016/01/04          &2457391	           &4.2           &1800          &82644390        &quiescence                    \\
CZ Ori                        &LAMOST J061643.21+152411.2        &0.2189               &UG        &GAC092N15M1          &2016/01/09          &2457396            &2.5           &1800          &82651570           &quiescence              \\
&        &               &        &GAC092N13B1          &2018/02/04          &2458153            &3.3           &1500          &83741539           &quiescence              \\
&        &               &        &HD060953N145109V01   &2018/10/10          &2458401            &2.5           &600          &84099144            &quiescence             \\
AY Psc                        &LAMOST J013655.46+071628.7        &0.217321             &ZC        &EG014125N052915M01   &2015/10/15          &2457310            &2.9           &1800          &82527873           &outburst peak              \\
&        &               &        &EG012903N071738M01   &2015/12/12          &2457368            &2.9           &1800          &82611051           &early decline              \\
&        &               &        &EG014125N052915B01   &2017/12/14          &2458101            &4.0           &1500          &83666644           &standstill              \\
SDSS J081610.83+453010.1      &LAMOST J081610.83+453010.1        &0.2096                &UG              &HD081449N430205M01   &2015/02/12          &2457065            &2.8           &1800          &82174979          &$-$                \\
TW Tri                        &LAMOST J013637.01+320040.0        &0.207584               &UG           &M31\underline{ }019N31\underline{ }B1        &2011/12/13          &2455908       &3.0    &900     &80541774   &early rise                    \\
&        &               &           &M31023N33B1          &2013/09/25          &2456560            &3.8           &1500          &81447897         &outburst peak                    \\
&        &               &           &M31023N33B2          &2013/09/25          &2456560            &3.8           &1500          &81447965         &outburst peak                    \\
&        &               &           &VB021N31V1           &2013/10/04          &2456569            &3.5           &600           &81460782         &late decline                    \\
&        &               &           &HD014422N323057B01   &2015/10/01          &2457296            &2.5           &1500          &82507715         &rise                    \\
SDSS J080846.19+313106.0      &LAMOST J080846.19+313106.0        &0.20587                &UG              &GAC\underline{ }122N29\underline{ }M1        &2012/01/26          &2455952     &$-$      &1200    &80572302      &quiescence                 \\
AT Cnc                        &LAMOST J082836.91+252003.0        &0.2011               &ZC        &GAC125N25V1	        &2013/04/14           &2456396            &3.6           &600           &81211445          &outburst peak                \\
SDSS J100658.41+233724.4      &LAMOST J100658.40+233724.4        &0.18591331                 &UG              &HD100153N235852M01   &2015/02/14          &2457067            &3.2           &1800          &82177920          &quiescence                \\
HH Cnc                        &LAMOST J091650.76+284943.1        &0.1845               &UG        &HD090800N293915M01   &2013/12/02          &2456628	           &3.5           &1800          &81545950           &quiescence              \\
&        &               &        &HD091735N272519M01   &2014/12/20          &2457011	           &4.8           &1800          &82097481           &quiescence              \\
&        &               &        &HD091735N272519M02   &2016/11/26          &2457718	           &2.3           &1800          &83115587           &quiescence              \\
SS Aur                        &LAMOST J061322.42+474425.1        &0.1828               &UG           &GAC094N47B1          &2017/03/02          &2457814            &2.5           &1500          &83253315         &quiescence                    \\
V405 Peg                      &LAMOST J230949.09+213517.0        &0.177647                &DN?              &EG231359N193802B01   &2015/11/29          &2457355            &3.2           &1500          &82592291          &quiescence                \\
U Gem                         &LAMOST J075505.23+220005.0        &0.176906               &UG           &GAC117N24B1          &2012/10/29          &2456229            &4.8           &1200          &80971515         &quiescence                    \\
&        &               &           &GAC117N24V1          &2012/12/04          &2456265            &3.0           &600           &81023274         &outburst peak                    \\
&        &               &           &GAC117N24V2          &2012/12/04          &2456265            &3.0           &600           &81023313         &outburst peak                    \\
GY Cnc                        &LAMOST J090950.53+184947.4        &0.175442               &UG        &HD090902N172810B01   &2013/02/02          &2456325            &3.5           &900           &81109419           &quiescence              \\
&        &               &        &HD090902N172810M01   &2013/02/02          &2456325            &3.3           &1200          &81109469           &quiescence              \\
X Leo                         &LAMOST J095101.46+115231.2        &0.1644                &UG              &HD095839N105737V01   &2016/11/27          &2457719            &2.3           &600           &83117132          &decline                \\
SDSS J094002.56+274942.0      &LAMOST J094002.56+274942.0        &0.16352                &UG              &F5591705             &2011/12/21          &2455916            &4.3           &1800          &80520639          &quiescence                \\
&        &                &              &HD094138N255446M01   &2015/12/12          &2457368            &2.7           &1800          &82611644          &quiescence                \\
&        &                &              &HD093318N282204M02   &2017/01/01          &2457754            &2.7           &1800          &83167288          &quiescence                \\
AR And                        &LAMOST J014503.27+375633.1        &0.1630               &UG        &M31025N38B2          &2016/01/05          &2457392            &4.8           &1500          &82645595           &rise              \\
SDSS J105550.08+095620.4      &LAMOST J105550.08+095620.4        &0.1624                &UG              &HD104915N103242M02   &2017/04/27          &2457870            &2.7           &1800          &83334023          &quiescence                \\
V367 Peg                     &LAMOST J224500.72+165513.3        &0.1619                &DN              &EG224702N171358M01   &2017/11/14          &2458071            &3.7           &1800         &83623383          &quiescence               \\
SDSS J081256.85+191157.8      &LAMOST J081256.85+191157.6        &0.16005248                &NL/DN             &GAC125N19B2          &2016/01/13          &2457400            &3.5           &1500          &82657514          &$-$                \\
IP Peg                        &LAMOST J232308.57+182459.8        &0.158206               &UG           &EG231359N193802V01   &2015/11/29          &2457355            &3.3           &600           &82592385         &quiescence                  \\
\hline
\end{tabular}
\end{center}
\end{table*}
\end{landscape}

\addtocounter{table}{-1}
\begin{landscape}
\begin{table*}
\caption{$-$continued.}
 \begin{center}
\tiny
 \begin{tabular}{lllllllllll}\hline\hline
 Targets                       &LAMOST name                      &$P_{orb}$(d)   &Type     &Planid               &Date             &MJD              &Seeing($''$)        &Exp.T(s)       &MJM       &States                  \\\hline
VZ Sex                        &LAMOST J094431.73+035805.4        &0.1487                &UG              &F5597602             &2012/02/18          &2455975            &5.4           &1800          &80605366          &quiescence                \\
&        &                &              &HD094623N014708B02   &2017/03/08          &2457820            &3.4           &1500          &83262107          &quiescence                \\
SDSS J213559.30+052700.5      &LAMOST J213559.30+052700.5        &0.137                 &NL/DN              &EG213118N034906B01   &2015/10/06          &2457301            &2.8           &1500          &82514650          &$-$                \\
V1239 Her                     &LAMOST J170213.25+322954.1        &0.100082                &SU              &HD165712N321400M01   &2013/04/13          &2456395            &4.6           &1200          &81210399          &quiescence                \\
&        &                &              &HD165712N321400M02   &2017/04/25          &2457868            &2.8           &1800          &83331490          &quiescence                \\
&        &                &              &HD171048N321044M02   &2017/05/23          &2457898            &3.0           &1800          &83371737          &quiescence                \\
NY Ser                        &LAMOST J151302.30+231508.6        &0.0978               &SU           &HD150715N223351B     &2014/04/21          &2456768            &3.6           &1500          &81747441         &quiescence                   \\
&        &               &           &HD150715N223351B02   &2016/03/06          &2457453            &3.1           &1500          &82733957         &quiescence                   \\
SDSS J153634.42+332851.9      &LAMOST J153634.42+332851.9        &0.0921                &SU              &HD153842N313306M01   &2015/05/12          &2457154            &3.8           &1800          &82303171          &quiescence                \\
GZ Cnc                        &LAMOST J091551.67+090049.5        &0.0881               &SU        &VB138N06V1           &2014/04/03          &2456750            &4.0           &600           &81721200           &quiescence              \\
&        &               &        &VB138N06V2           &2014/04/03          &2456750            &3.9           &600           &81721256           &quiescence              \\
V344 Lyr                      &LAMOST J184439.18+432227.8        &0.087904               &SU           &HD184435N434959B01   &2016/05/07          &2457515            &3.5           &1500          &82823188         &outburst peak?                    \\
SDSS J113950.58+455817.9      &LAMOST J113950.58+455817.9        &0.0843                &SU              &HD113019N471440F01   &2013/04/02          &2456384            &2.9           &1800          &81194250          &$-$                \\
SDSS J155720.75+180720.2      &LAMOST J155720.75+180720.2        &0.081                &SU              &HD160349N174809M01   &2017/03/27          &2457839            &2.5           &1800          &83289804          &quiescence                \\
AC LMi                        &LAMOST J101947.26+335753.6        &0.0792               &SU        &HD102408N334306M01   &2014/12/25          &2457016            &2.5           &1800          &82104698           &quiescence              \\
                                                                 & & &        &HD102408N334306M02   &2017/02/26          &2457810            &2.2           &1800          &83247846           &quiescence              \\
V660 Her                      &LAMOST J174209.16+234829.6        &0.07826                &SU              &HD174640N254456F01   &2013/06/03          &2456446            &3.6           &1800          &81283693          &quiescence                \\
HY Psc                        &LAMOST J230351.64+010651.0        &0.0767               &SU           &EG230517N011825M01   &2015/10/08          &2457303            &3.6           &1800          &82517596         &quiescence                  \\
WY Tri                        &LAMOST J022500.46+325955.5        &0.0759                &SU              &GACII036N32M1        &2017/11/13          &2458070            &3.2           &1800          &83622130          &quiescence                \\
HT Cas                        &LAMOST J011013.16+600435.2        &0.073647               &SU           &GACII018N60B1        &2017/12/17          &2458104            &3.2           &1500          &83670895         &quiescence                  \\
CC Cnc                        &LAMOST J083619.14+212105.3        &0.07352               &SU        &HD084620N203659M02   &2018/01/17          &2458135            &2.2           &1800          &83715888           &quiescence              \\
CY UMa                        &LAMOST J105656.98+494118.3        &0.06957               &SU        &F5591506             &2011/12/19          &2455914            &2.6           &1800          &80517892           &late outburst              \\
&        &               &        &HD105331N491352M01   &2013/04/28          &2456410            &3.1           &1200          &81231655           &outburst end              \\
&        &               &        &HD105331N491352M02   &2018/01/12          &2458130            &2.3           &1800          &83708825           &quiescence              \\
SDSS J083845.23+491055.5      &LAMOST J083845.23+491055.4        &0.0692                &SU              &HD082820N494114M01   &2015/01/23          &2457045            &2.7           &1800          &82146248          &quiescence                \\
&        &                &              &HD082820N494114M02   &2017/12/11          &2458098            &3.3           &1800          &83662730          &quiescence                \\
V1208 Tau                     &LAMOST J045944.03+192622.6        &0.0687                &SU              &GAC073N19B1          &2014/10/06          &2456936            &3.2           &1500          &81989445          &quiescence                \\
V368 Peg                      &LAMOST J225843.48+110911.9        &0.0686                &SU              &EG225624N130316B01   &2015/10/03          &2457298            &3.4           &1500          &82510395          &quiescence                \\
IR Gem                        &LAMOST J064734.72+280622.1        &0.0684               &SU           &GAC\underline{ }101N28\underline{ }B1        &2011/11/09          &2455875     &2.2    &900    &80460203      &quiescence                   \\
&     &        &                                   &GAC\underline{ }100N28\underline{ }B1        &2012/01/12          &2455938     &3.0    &900    &80552106      &quiescence                   \\
TY Psc                        &LAMOST J012539.36+322308.5        &0.06833               &SU           &M31\underline{ }021N31\underline{ }B2        &2011/11/10          &2455875     &4.0       &1200    &80461434  &$-$                 \\
KS UMa                        &LAMOST J102026.52+530433.0        &0.0680               &SU           &HD100510N522216M01   &2015/02/18          &2457071            &2.6           &1800          &82183636         &quiescence                   \\
SX LMi                        &LAMOST J105430.43+300610.1        &0.0672               &SU           &B90706               &2011/12/11          &2455906            &3.6           &900           &80506392         &quiescence                    \\
&        &               &           &HD104953N275826M01   &2014/01/27          &2456684            &4.2           &2400          &81626505         &quiescence                    \\
&        &               &           &HD105033N320139M02   &2017/01/01          &2457754            &3.1           &1800          &83167409         &quiescence                    \\
&        &               &           &HD104953N275826M02   &2018/01/14          &2458132            &3.5           &1800          &83711649         &quiescence                    \\
US 691                        &LAMOST J093249.57+472522.8        &0.06630354               &SU           &HD094225N484633M01   &2016/01/09          &2457396            &3.7           &1800          &82651830         &quiescence                    \\
&        &               &           &HD091915N465637M01   &2016/12/22          &2457744            &3.4           &1800          &83152935         &quiescence                    \\
&        &               &           &HD091915N465637M01   &2016/12/26          &2457748            &4.1           &1800          &83158704         &quiescence                    \\
GO Com                        &LAMOST J125637.10+263643.2        &0.0658               &SU        &HD125932N280356F01   &2013/04/13          &2456395            &3.6           &1800          &81210157           &near rise              \\
&        &               &        &HD125932N280356M02   &2017/04/25          &2457868            &2.0           &1800          &83331231           &quiescence              \\
2MASS J04260931+3541451      &LAMOST J042609.33+354144.8        &0.06560                &SU              &GAC069N36B1          &2013/10/31          &2456596            &3.1           &1500          &81499750          &quiescence       \\
&        &                &              &GAC066N34B1          &2014/01/09          &2456666            &4.0           &1500          &81600253          &quiescence                \\
&        &                &              &GAC066N34B2          &2015/12/31          &2457387            &4.6           &1500          &82638511          &quiescence                \\
&        &                &              &GAC069N36B2          &2016/02/08          &2457427            &2.4           &1500          &82694560          &quiescence                \\
AK Cnc                        &LAMOST J085521.18+111815.3        &0.0651               &SU        &KP084656N120635B01   &2016/02/06          &2457424            &3.1           &1500          &82692010           &quiescence              \\
HW Boo                        &LAMOST J134323.16+150916.8        &0.06435               &DN           &HD133420N140639M02   &2018/01/17          &2458135            &2.5           &1800          &83716125         &quiescence                  \\
ER UMa                        &LAMOST J094711.94+515408.9        &0.06366               &SU        &HD093959N532949B02   &2016/04/12          &2457490            &3.3           &1500          &82786813           &quiescence              \\
&        &               &        &HD093959N532949V01   &2016/12/01          &2457723            &3.3           &600           &83122913           &early rise              \\
MR UMa                        &LAMOST J113122.39+432238.5        &0.0633               &SU           &HD113736N444258B01   &2013/05/14          &2456426            &3.8           &600           &81254705         &quiescence                   \\
DW Cnc                        &LAMOST J075853.03+161645.1        &0.059793               &NL/IP        &F5591504             &2011/12/19          &2455914            &2.5           &1800          &80517720           &quiescence              \\
&        &               &        &GAC120N18B1          &2014/11/04          &2456965            &3.1           &1500          &82031273           &quiescence              \\
&        &               &        &GAC120N14B2          &2015/12/21          &2457377            &3.0           &1500          &82624485           &quiescence              \\
&        &               &        &GAC120N14B1          &2015/12/21          &2457377            &2.2           &1500          &82624390           &quiescence              \\
\hline
\end{tabular}
\end{center}
\end{table*}
\end{landscape}

\addtocounter{table}{-1}
\begin{landscape}
\begin{table*}
\caption{$-$continued.}
 \begin{center}
\tiny
 \begin{tabular}{lllllllllll}\hline\hline
 Targets                       &LAMOST name                      &$P_{orb}$(d)   &Type     &Planid               &Date             &MJD              &Seeing($''$)        &Exp.T(s)       &MJM       &States                  \\\hline
FS Aur                        &LAMOST J054748.36+283511.2        &0.059581               &SU        &GAC\underline{ }089N28\underline{ }B2    &2011/11/10          &2455875       &2.0     &900    &80461647         &late decline              \\
&        &               &        &GAC089N28B1          &2013/03/06          &2456357            &3.6           &1200          &81155228           &quiescence              \\
EZ Lyn                        &LAMOST J080434.19+510349.4        &0.059005               &SU        &HD081044N520834M01   &2013/01/08          &2456300            &4.5           &1200          &81073547           &after superoutburst              \\
NZ Boo                        &LAMOST J150240.98+333423.8        &0.058910               &SU           &HD150818N334223M01   &2014/05/22          &2456800            &3.7           &1800          &81791906         &quiescence                   \\
EG Cnc                        &LAMOST J084303.98+275149.6        &0.05877               &SU        &HD084531N282643F01   &2013/01/15          &2456307            &2.9           &1800          &81083529           &quiescence              \\
&        &               &        &HD084531N282643M01   &2014/12/24          &2457015	   	       &3.7           &1800          &82103174           &quiescence              \\
&        &               &        &HD084531N282643M02   &2016/11/24          &2457716            &2.3           &1800          &83112739           &quiescence              \\
RZ LMi                        &LAMOST J095148.96+340723.5        &0.0584               &SU           &HD095000N333605B01   &2013/11/12          &2456608            &$-$            &1500          &81517297          &outburst peak                   \\
V355 UMa                      &LAMOST J133941.12+484727.4        &0.05731               &SU           &HD132901N475049M01   &2013/04/11          &2456393            &2.4           &1200          &81207358         &quiescence                    \\
        & &  &                              &HD132901N475049M02   &2017/04/27          &2457870            &2.6           &1800          &83334146          &quiescence               \\
PQ And                        &LAMOST J022929.60+400239.9        &0.0560               &SU           &GACII034N39M1        &2017/09/22          &2458018            &2.7           &1800          &83547441         &quiescence                   \\
V627 Peg                &LAMOST J213806.55+261958.1        &0.05452               &SU        &HD213922N242940B01   &2016/10/03          &2457664            &2.9           &1500          &83037397           &early rise              \\
EI Psc                        &LAMOST J232954.18+062812.0        &0.044567               &SU        &EG233013N050046B01   &2013/09/25          &2456560            &3.8           &1500          &81447765           &early rise              \\
YZ LMi                        &LAMOST J092638.71+362402.4        &0.019661                &AC/DN              &HD093539N354836M01   &2016/01/07          &2457394            &4.2           &1800          &82648923          &quiescence                \\
&        &                &              &HD091850N364809M01   &2016/12/24          &2457746            &3.2           &1800          &83155782          &quiescence                \\
PTF1 J071912.13+485834.0      &LAMOST J071912.12+485834.5        &0.01859               &AC/BWD           &GAC112N47M1          &2016/12/19          &2457741            &2.4           &1800          &83148524         &outburst                   \\
UCAC3 232-65219               &LAMOST J063213.00+253622.5        &               &UG           &KP063626N271642V01   &2016/01/18          &2457405            &6.3           &600           &82664615         &quiescence                    \\
SDSS J003303.94+380105.4      &LAMOST J003303.94+380105.4        &                &UG              &M31007N36M2          &2015/10/16          &2457311            &3.3           &1800          &82529212          &quiescence                \\
&        &                &              &M31005N38M1          &2015/10/17          &2457312            &2.9           &1800          &82530637          &quiescence                \\
SDSS J084358.08+425036.8      &LAMOST J084358.08+425036.8        &                &UG              &HD084415N420602M01   &2016/02/08          &2457426            &3.3           &1800          &82694784          &quiescence                \\
SDSS J090628.24+052656.9      &LAMOST J090628.25+052656.9        &                &UG?              &HD091212N035201M01   &2016/03/05          &2457452            &2.4           &1800          &82732202          &rise?                \\
SDSS J124417.89+300401.0      &LAMOST J124417.89+300401.0        &                &UG              &HD124107N302613M02   &2017/02/01          &2457785            &3.3           &1800          &83212004          &quiescence                \\
SDSS J153015.04+094946.3      &LAMOST J153015.04+094946.3        &                &SU              &HD153553N111556M01   &2017/04/25          &2457868            &2.6           &1800          &83331362          &quiescence                \\
\noalign{\smallskip}\hline
\end{tabular}
\end{center}
\end{table*}
\end{landscape}

\begin{figure}
\begin{center}
\includegraphics[width=1.0\columnwidth]{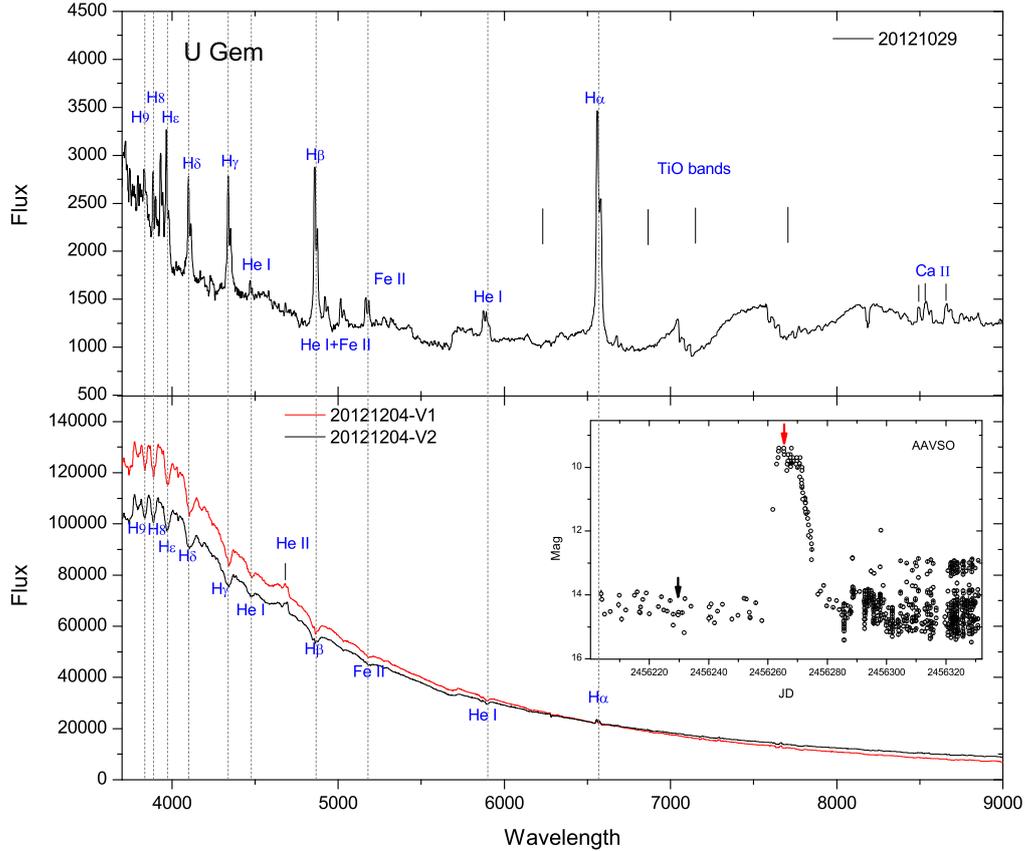}
\caption{LAMOST spectra of U Gem were observed in quiescence and at the outburst peak. The quiescent spectrum in the upper panel shows prominent disc emission together with TiO absorption bands from the cool M-type donor. The bottom panel and the inset display the outburst spectrum and the AAVSO light curve, respectively. Two arrows correspond to the observed time of the spectra.}
\end{center}
\end{figure}

\begin{figure}
\begin{center}
\includegraphics[width=1.0\columnwidth]{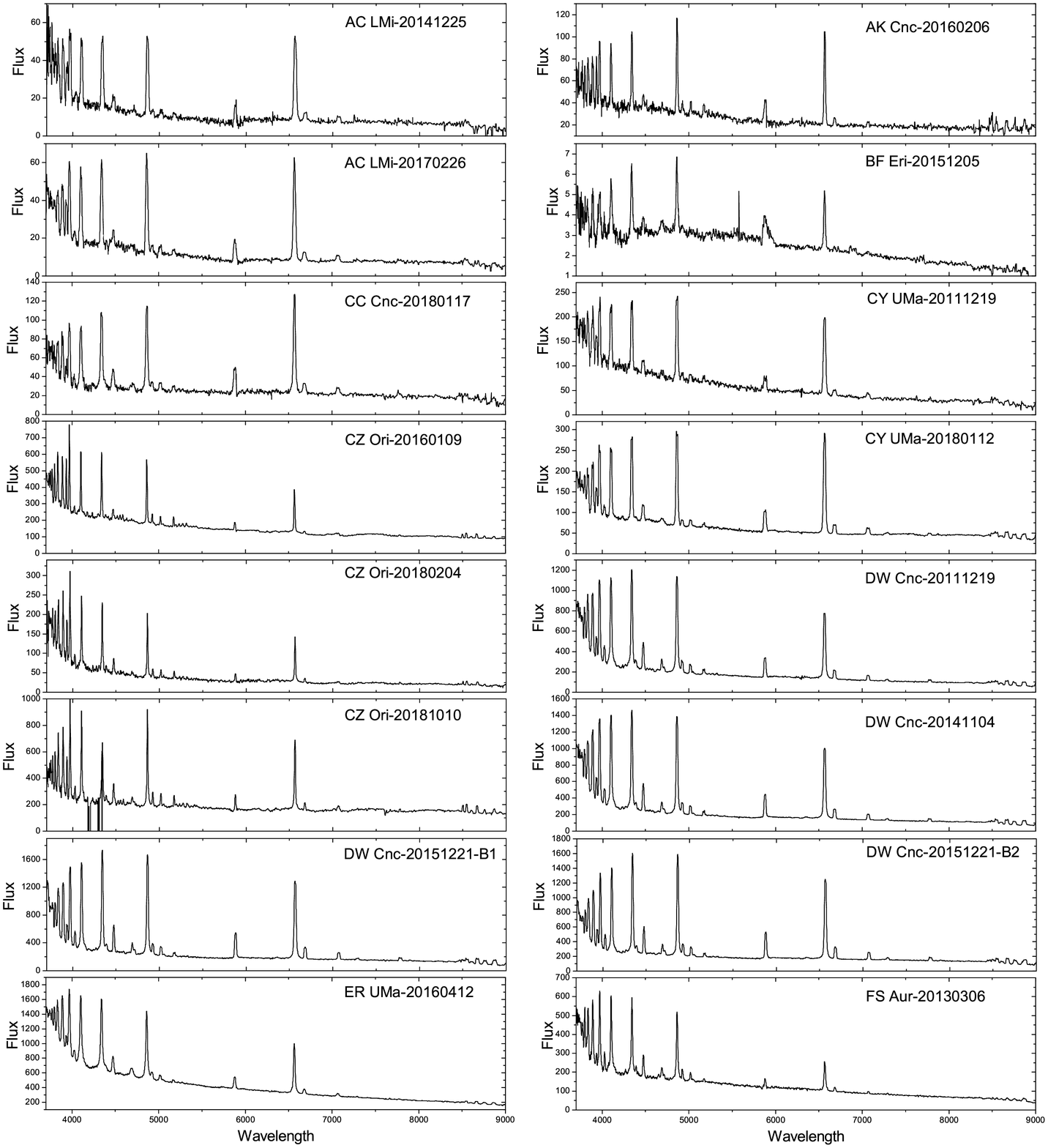}
\caption{LAMOST quiescence spectra of nine dwarf novae: AC LMi, AK Cnc, BF Eri, CC Cnc, CY UMa, CZ Ori, DW Cnc, ER UMa and FS Aur.}
\end{center}
\end{figure}

\begin{figure}
\begin{center}
\includegraphics[width=1.0\columnwidth]{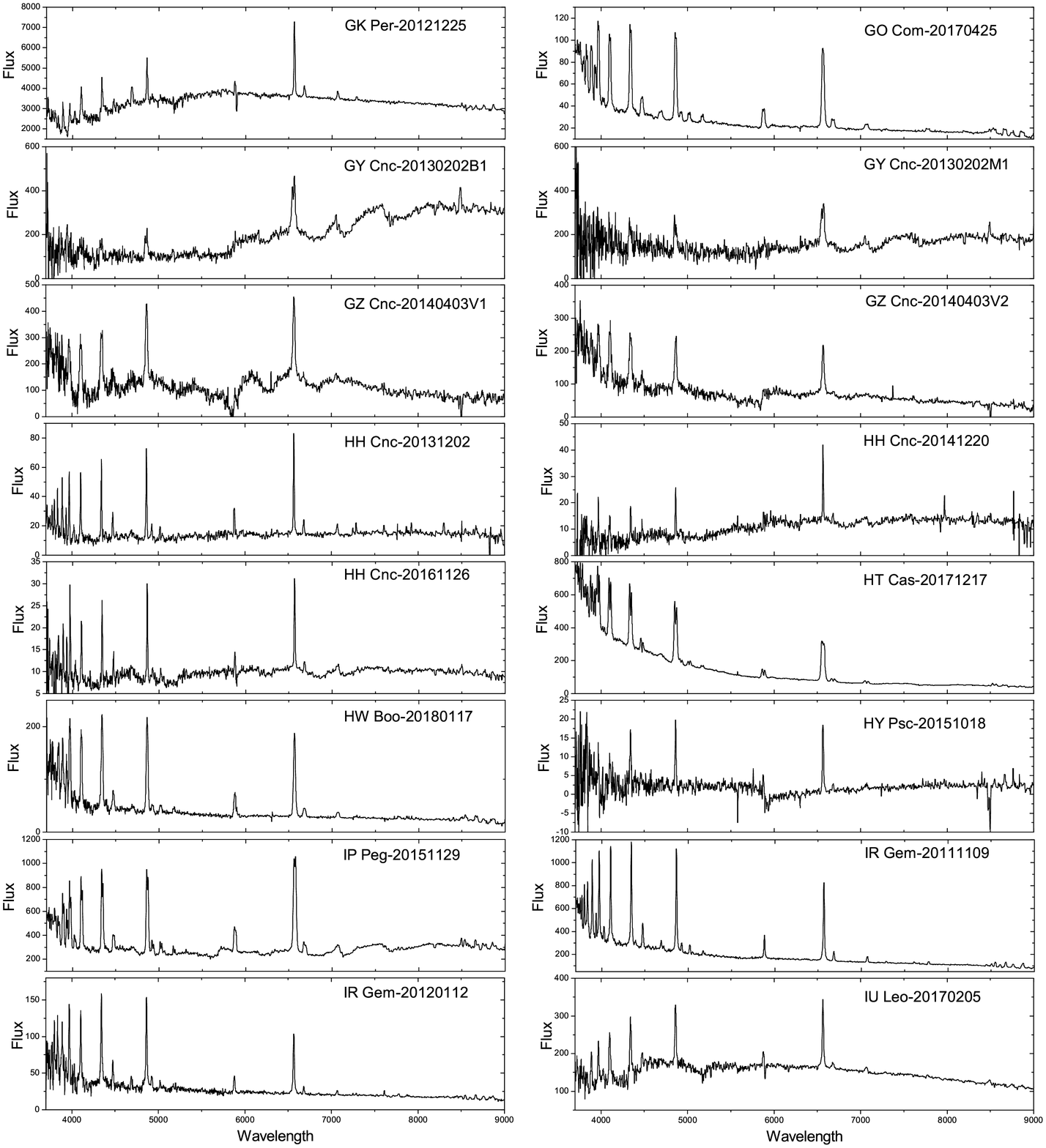}
\caption{LAMOST quiescence spectra of eleven dwarf novae: GK Per, GO Com, GY Cnc, CZ Cnc, HH Cnc, HT Cas, HW Boo, HY Psc, IP Peg, IR Gem and IU Leo. }
\end{center}
\end{figure}

\begin{figure}
\begin{center}
\includegraphics[width=1.0\columnwidth]{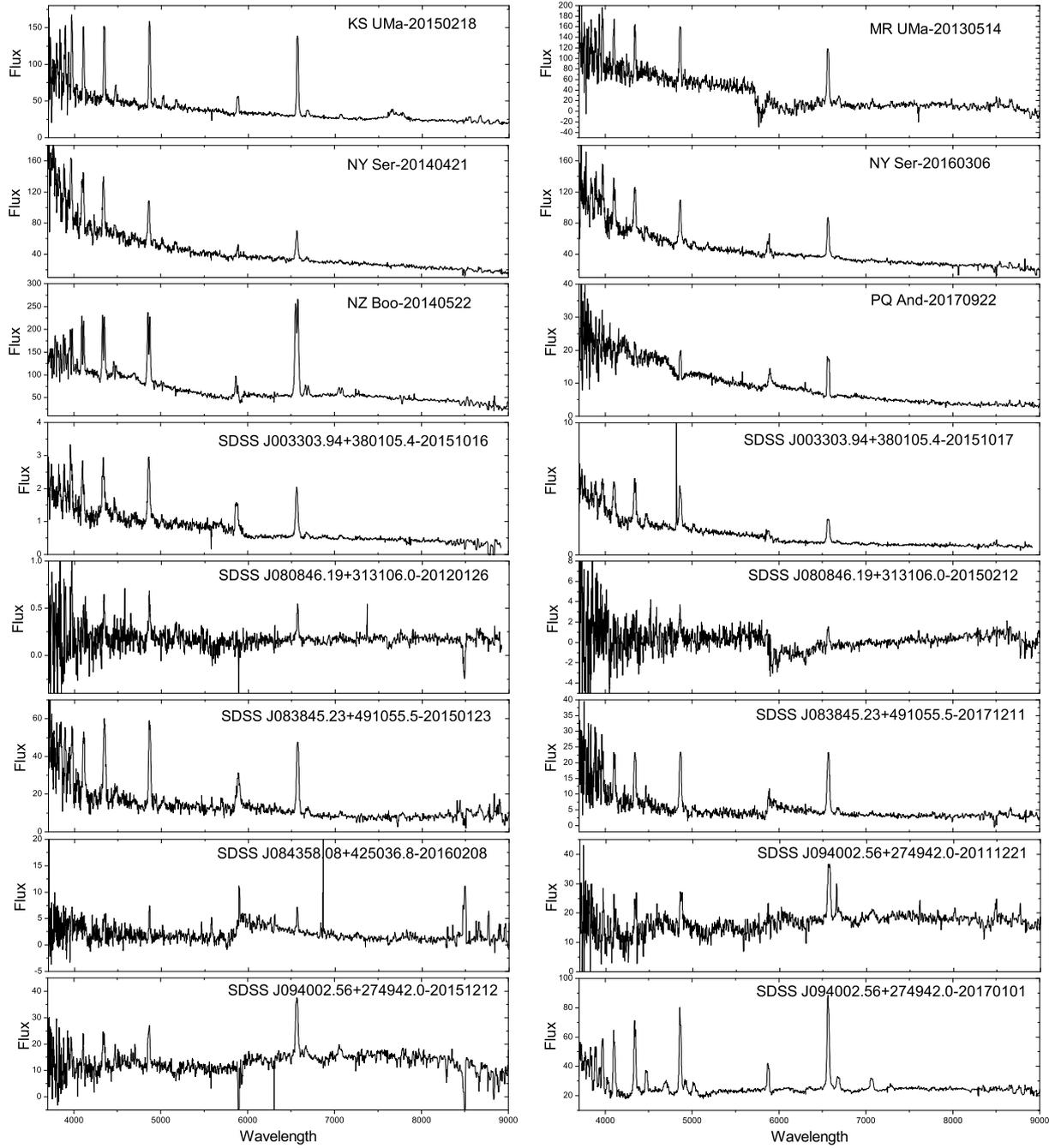}
\caption{LAMOST quiescence spectra of ten dwarf novae: KS UMa, MR UMa, NY Ser, NZ Boo, PQ And, SDSS J003303.94+380105.4, SDSS J080846.19+313106.0, SDSS J083845.23+491055.5, SDSS J084358.08+425036.8 and SDSS J094002.56+274942.0.}
\end{center}
\end{figure}

\begin{figure}
\begin{center}
\includegraphics[width=1.0\columnwidth]{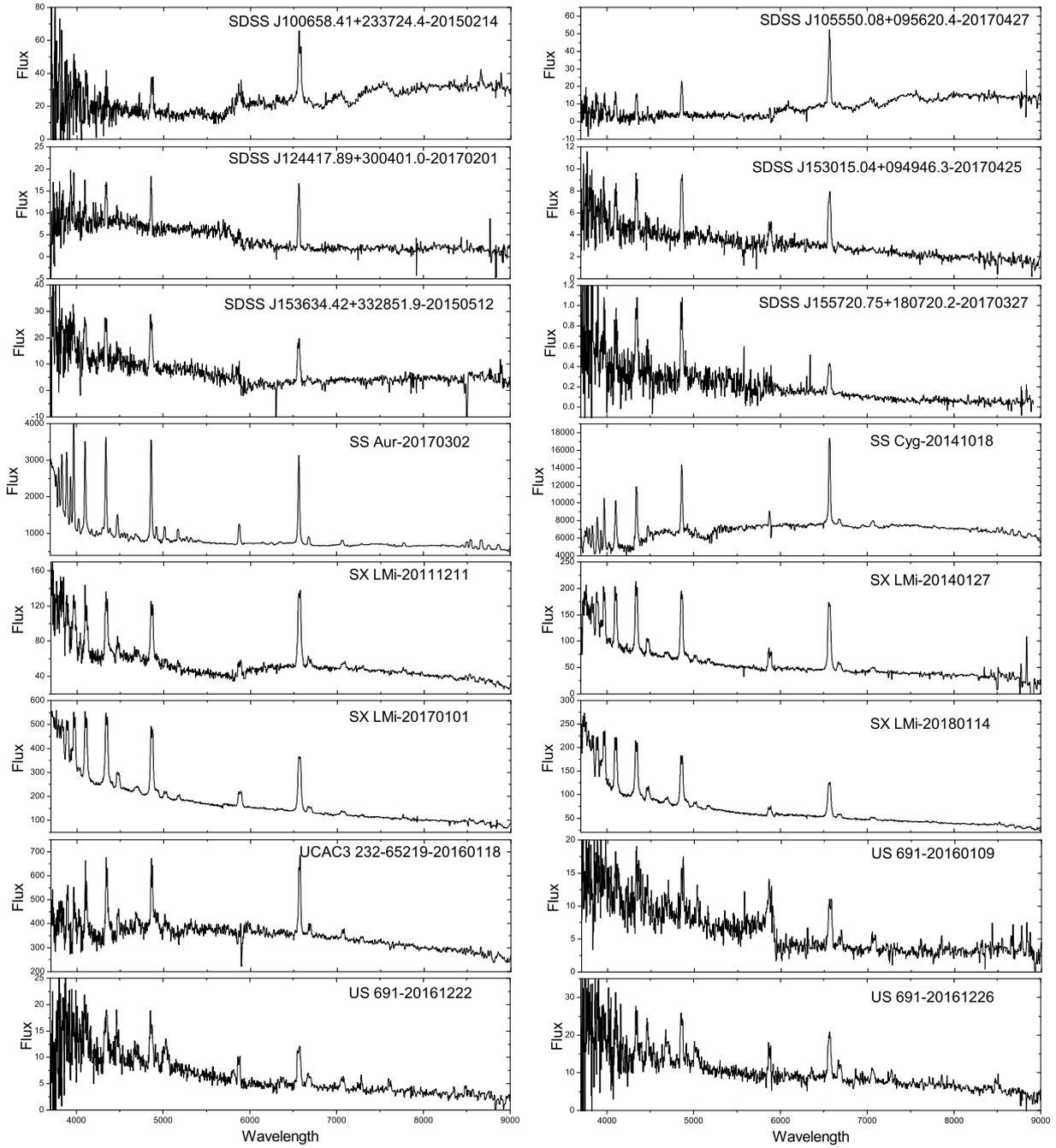}
\caption{LAMOST quiescence spectra of eleven dwarf novae: SDSS J100658.41+233724.4, SDSS J105550.08+095620.4, SDSS J124417.89+300401.0, SDSS J153015.04+094946.3, SDSS J153634.42+332851.9, SDSS J155720.75+180720.2, SS Aur, SS Cyg, SX LMi, UCAC3 232-65219 and US 691.}
\end{center}
\end{figure}

\begin{figure}
\begin{center}
\includegraphics[width=1.0\columnwidth]{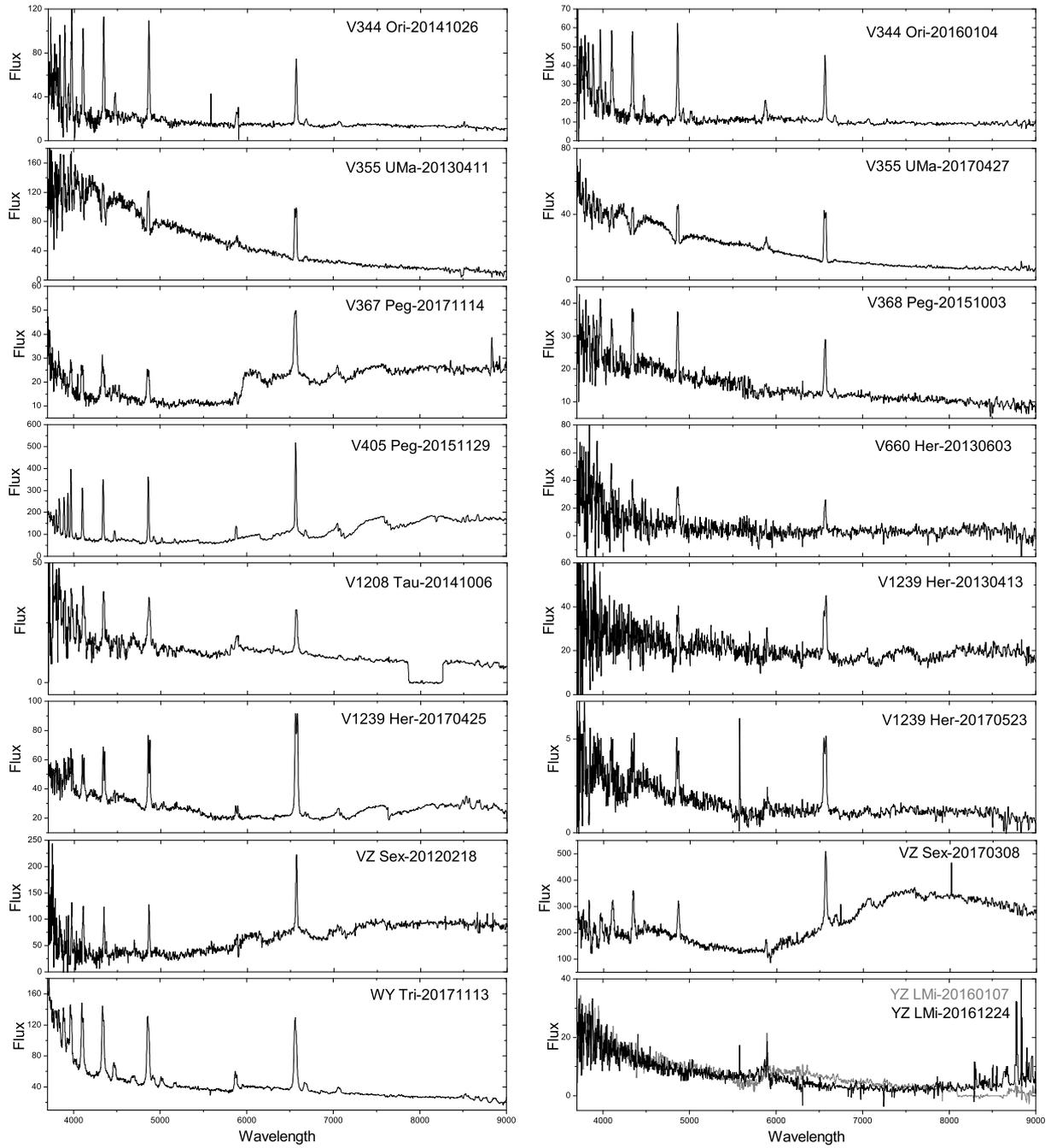}
\caption{LAMOST quiescence spectra of eleven dwarf novae: V344 Ori, V355 UMa, V367 Peg, V368 Peg, V405 Peg, V660 Her, V1208 Tau, V1239 Her, VZ Sex, WY Tri and YZ LMi (AM CVn star).}
\end{center}
\end{figure}

\begin{figure}
\begin{center}
\includegraphics[width=1.0\columnwidth]{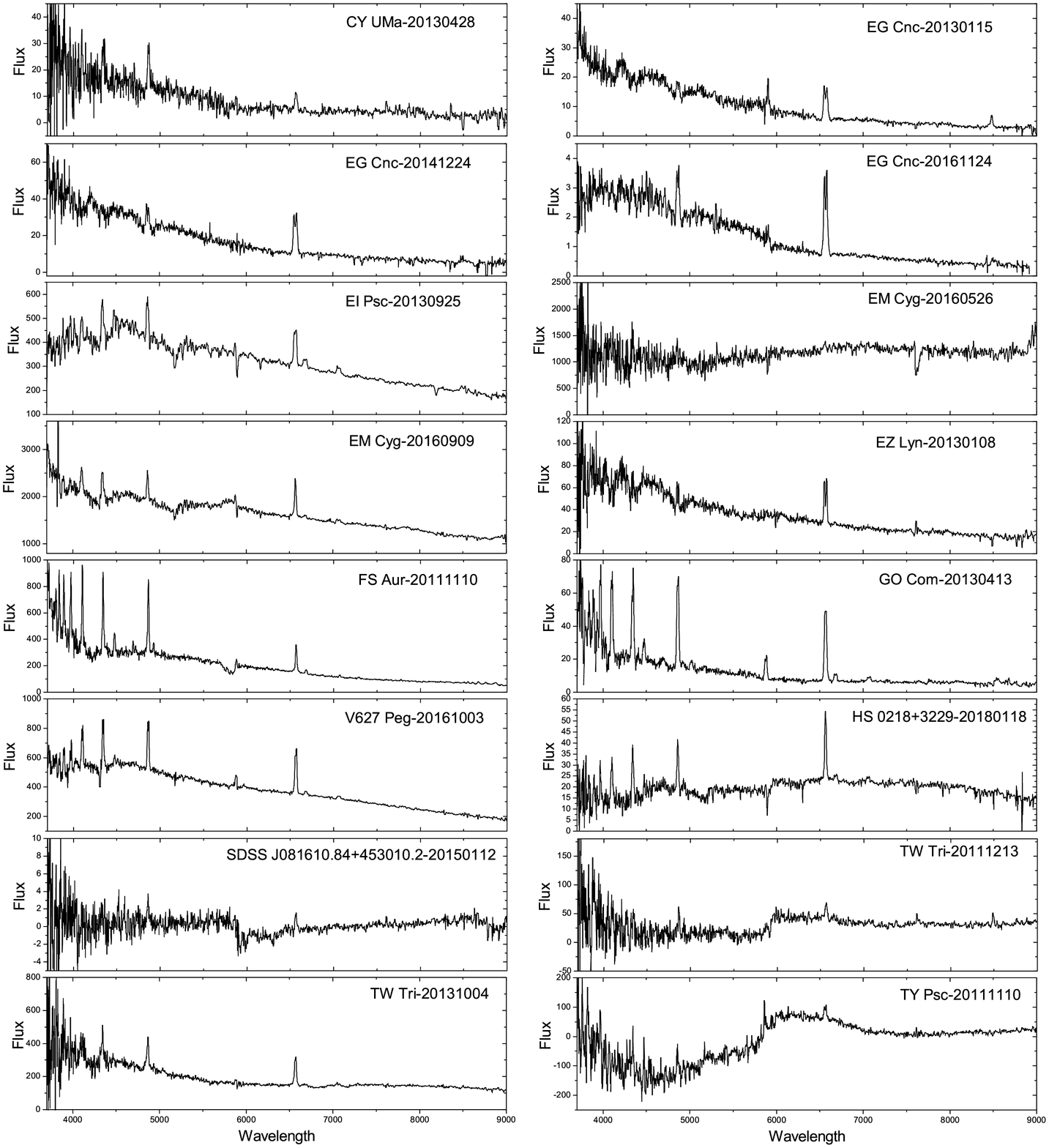}
\caption{LAMOST spectra obtained near outburst (before and after) of twelve dwarf novae: CY UMa, EG Cnc, EI Psc, EM Cyg, EZ Lyn, FS Aur, GO Com, V627 Peg, HS 0218+3229, SDSS J081610.84+453010.2, TW Tri and TY Psc.}
\end{center}
\end{figure}

\emph{V367 Peg}. The LAMOST has observed the first spectrum for this object on 2017 Nov 14. This is a typical quiescent spectrum. Even though there is a data-reduction issue around $\lambda$6000{\AA}, the M-type donor feature is visible clearly. Combined with its orbital period of 3.89 h (Woudt et al. 2005), we estimated the donor's spectral type as M3-5.

\emph{VZ Sex}. This object was observed in quiescence on 2012 Feb 18 and 2017 March 8, respectively. As noted above, its spectra are dominated by the continuum from an M-type donor. The donor's spectral type was estimated to be M1-2 (Mennickent et al. 2002; Thorstensen et al. 2010). However, the shape of the continuums, especially spectrum taken on 2017 March 8, is unusual. The LAMOST spectrum on 2012 Feb 18 is similar to the SDSS spectrum in the database\footnote{http://skyserver.sdss.org/dr13/en/tools/quicklook/summary.aspx?id=1237654601561735170}, which shows a redder continuum. This feature looks like a K star. The Na D absorption lines are also seen in these spectra. We therefore suggest that the donor may be a late K star.

\emph{YZ LMi(=SDSS J092638.71+362402.4)}. Two unusual spectra obtained on 2016 Jan 7 and Dec 24 show the blue continuum and several helium emission lines, typical of AM CVn binaries in quiescence. This object has been confirmed to be an eclipsing AM CVn star, with an orbital period of 28.3 min (Anderson et al. 2005). Compared with the SDSS spectrum (Anderson et al. 2005), the LAMOST spectra show a strong HeI 5875 line, but no trace of HeII 4686 line. However, the spectral feature of the eclipsing systems, broad double-peaked emission lines, is only seen in HeI 5875 due to low signal-to-noise ratio.

In fact, the unusual features near $\lambda$6000{\AA} and  $\lambda$8500{\AA} also appear in other targets except for the above mentioned samples, they are: GZ Cnc, MR UMa, SDSS J080846.19+313106.0, SDSS J081610.84+453010.2, SDSS J153634.42+332851.9, UCAC3 232-65219 and US 691. We suggest that all peculiar behaviors around these two wavelengths in such spectra should be attributed to the known bad regions. In Fig. 7, moreover, several interesting spectra near outburst are notable. In EI Psc, EM Cyg (2016 Sept 9), GO Com, V627 Peg and TW Tri (2011 Dec 13), the spectra were obtained before outburst, by checking the AAVSO light curves. These spectra are similar to the quiescent spectra, but have a bluer continuum. The spectra of other several targets were carried out only a few days after an outburst, including CY UMa (2013 April 28), EM Cyg (2016 May 26), FS Aur, TW Tri (2013 Oct 4). These spectra show a continuum appearance similar to the quiescent spectra, but the H$\alpha$ line is weak or absence. Due to the lack of AAVSO data, the activity state of two unusual spectra, HS 0218+3229 on 2018 Jan 18 and TY Psc on 2011 Nov 10, was not determined. The former shows a similar appearance to its quiescent spectrum, but with some special absorption lines at $\lambda$5894, 6302, 8350{\AA}. The latter displays an exceptional shape of the red continuum and the negative flux in the blue channel. Also, there is a sudden change in the continuum near $\lambda$6000{\AA}.

\begin{table*}
\tiny
\centering
\caption{Equivalent widths (EWs) and the full width at halfmaximum (FWHM) of LAMOST spectral lines.}
\label{Tab03}
\begin{tabular}{lccccccccc}
\toprule
\multirow{4}{*}{Target} & \multirow{4}{*}{Date-Obs}   & \multicolumn{2}{c}{H$\alpha$} & \multicolumn{2}{c}{H$\beta$} & \multicolumn{2}{c}{H$\gamma$}  & \multicolumn{2}{c}{HeII 4686} \\
\cmidrule(r){3-4} \cmidrule(r){5-6} \cmidrule(r){7-8}  \cmidrule(r){9-10}
&
&  $EW$      &  $FWHM$

&  $EW$      &  $FWHM$

&  $EW$      &  $FWHM$

&  $EW$      &  $FWHM$   \\
\midrule
GK Per             &2012/12/25       & -17.18             & 14.68          & -13.11           & 17.42           & -11.26          & 18.04           & -           & -          \\
V537 Peg           &2014/11/11       & -10.69             & 8.00          & -4.47           & 5.27           & -2.30          & 4.32           & -0.03           & 0.74          \\
SY Cnc                     &2017/12/18       & -7.76        & 12.30          & -4.11           & 9.80           & -2.91          & 9.02           & -0.53           & 12.87          \\
                           &2017/12/21       & -10.62       & 12.21          & -8.87          & 11.56           & -8.88          & 12.15          & -0.87           & 18.03          \\
                           &2018/01/08       & -4.87        & 10.38          & -3.18          & 10.05           & -1.92          & 8.44           & -0.49           & 8.99               \\
                           &2018/01/10       & -7.56        & 10.78               &  -3.86               & 8.85                & -2.41               & 7.52                & -1.26                & 20.3               \\
                           &2018/01/11       & -7.05        & 11.18               &  -3.57               & 9.05                & -2.24               & 9.19                & -0.95                & 15.97               \\
                           &2018/01/12       & -10.56       & 10.97               & -5.30                & 8.43                 & -4.16               & 7.80                & -0.84                 & 15.21                \\
                           &2018/12/03       & -3.76        & 9.56               & -1.43                & 7.08                & -1.00               & 6.53                & -0.70                & 21.21                \\
IU Leo                     &2015/01/01       & -7.90        & 15.28            & -3.42      & 12.08            & -2.20        & 12.43         &  -       &  -                      \\
                           &2017/02/05       & -21.53      & 17.89         & -17.56            & 17.47        & -21.21         & 18.98        & -        &  -                \\
HS 0218+3229               &2018/01/18       & -30.46      & 21.10     & -30.82       & 21.7             & -23.41        & 17.91         & -       & -                      \\
EM Cyg                     &2016/05/26       &  -     & -     & -      & -           & -        & -         & -        & -                       \\
                           &2016/09/09       & -13.01      & 23.45      & -6.96      & 20.57            & -7.59        & 23.4         & -       & -                      \\
SS Cyg                     &2014/10/18       & -30.98      & 20.88     & -27.14      & 20.49            & -24.71        & 19.66         & -        & -                       \\
BF Eri                     &2015/12/05       & -30.01      & 16.10     & -20.43      & 17.44            & -29.20       & 23.27         & -        & -                       \\
V344 Ori                   &2014/10/26       & -84.76      & 19.7     & -91.97      & 17.32            & -68.35        & 16.82         & -        & -                       \\
                           &2016/01/04       & -70.34      & 20.48     & -92.11      & 18.25            & -89.59        & 19.12         & -        & -                       \\
CZ Ori                     &2016/01/09       & -32.02      & 13.6     & -29.66      & 12.54            & -23.27        & 11.96         &  -       & -                       \\
                           &2018/02/04       & -61.70      & 13.08     & -54.14      & 12.29            & -41.37        & 11.70         & -        & -                       \\
                           &2018/10/10       & -53.97      & 14.54     & -43.01      & 10.75            & -34.76        & 12.41         & -        & -                       \\
AY Psc                     &2015/10/15       & -12.84      & 24.00     & -3.83      & 18.6            & -1.25        & 7.36         & -4.83        & 26.58                      \\
                           &2015/12/12       & -15.62      & 24.26     & -5.98      & 19.55            & -3.18        & 18.9         & -4.45        & 22.43                       \\
                           &2017/12/14       & -22.40      & 26.25     & -7.25      & 23.24            & -        & -         & -        & -                       \\
SDSS J081610.83+453010.1   &2015/02/12       & -      & -      & -      & -            & -        & -         & -        &  -                      \\
TW Tri                     &2011/12/13       & -22.76      & 30.60      & -      & -            & -        & -         & -        & -                       \\
                           &2013/09/25       & -2.59      & 16.11      & 3.32      & 42.54            & 4.36        & 37.39         & -0.57        & 13.93                       \\
                           &2013/09/25       & -2.91      & 16.02     & 2.89      & 46.19            & 3.31        & 35.4         & -0.78        & 14.68                       \\
                           &2013/10/04       & -38.08      & 28.03     & -20.10      & 24.98            & -26.04        & 33.36         & -        & -                       \\
                           &2015/10/01       & -17.64      & 18.31     & -5.45      & 13.08            & -2.82        & 11.24         & -        & -                       \\
SDSS J080846.19+313106.0   &2012/01/26       & -40.21      & 19.02     & -23.44      & 22.00            & -30.28        & 21.13         & -        & -                       \\
AT Cnc                     &2013/04/14       & -5.39      & 3.65     & -      & -            & -        & -         & -        & -                       \\
SDSS J100658.41+233724.4   &2015/02/14       & -62.33      & 37.52     & -45.45      & 33.74            & -        & -         & -        & -                       \\
HH Cnc                     &2013/12/02       & -61.91      & 11.40     & -50.88      & 10.26            & -49.20        & 10.46         & -        & -                       \\
                           &2014/12/20       & -26.54      & 9.49     & -20.75      & 9.01            & -        & -         & -        & -                       \\
                           &2016/11/26       & -29.24      & 12.10     & -30.82      & 10.47            & -30.35        & 10.84         & -        & -                       \\
SS Aur                     &2017/03/02       & -71.36      & 19.69     & -67.36      & 18.24            & -55.11        & 18.12         & -        & -                       \\
V405 Peg                   &2015/11/29       & -87.45      & 17.64     & -74.08      & 14.65             & -69.28        & 14.33         & -        & -                       \\
U Gem                      &2012/10/29       & -73.29      & 35.99     & -34.53      & 28.66            & -18.40        & 28.09         & -        & -                      \\
                           &2012/12/04       & -2.57      & 37.29     & 2.24      & 35.48            & 3.24        & 37.11         & -1.30        & 28.57                       \\
                           &2012/12/04       & -2.85      & 35.43     & 2.63      & 43.96            & 2.41        & 34.29         & -1.29        & 25.89                       \\
GY Cnc                     &2013/02/02       & -62.06      & 52.41     & -59.00      & 44.63            & -28.69        & 35.46         & -        & -                       \\
                           &2013/02/02       & -52.54      & 45.31     & -34.69      & 35.5            & -        & -         & -        & -                       \\
X Leo                      &2016/11/27       & -15.39      & 17.54     & -6.77      & 13.86            & -3.74        & 12.12         & -        & -                       \\
SDSS J094002.56+274942.0   &2011/12/21       & -40.53      & 34.29     & -29.65      & 31.60            & -        & -         & -        & -                       \\
                           &2015/12/12       & -50.61      & 31.99     & -35.02      & 23.93            & -27.01        & 25.25         & -        & -                       \\
                           &2017/01/01       & -76.07      & 27.93     & -68.87      & 25.96            & -65.94        & 26.37         & -        & -                       \\
AR And                     &2016/01/05       & -2.60      & 9.76      & 5.35      & 34.56            & 4.87        & 25.99         & -0.98        & 44.13                       \\
SDSS J105550.08+095620.4   &2017/04/27       & -122.80      & 26.03     & -107.60      & 22.17            & -50.83         & 19.39         & -        & -                       \\
V367 Peg                   &2017/11/14       & -47.73      & 36.90     & -44.02      & 35.22            & -14.80        & 20.73         &  -       & -                       \\
SDSS J081256.85+191157.8   &2016/01/13       & -35.42      & 20.85     & -10.83      & 15.42            & -5.25        & 13.70         & -1.70       & 29.40                     \\
IP Peg                     &2015/11/29       & -119.60      & 36.29     & -85.79      & 29.37            & -69.03        & 27.24         & -        & -                       \\
VZ Sex                     &2012/02/18       & -35.70      & 17.73     & -29.94      & 13.46            & -38.29        & 15.07         & -        & -                       \\
                           &2017/03/08       & -32.43      & 24.42     & -17.65      & 23.29            & -20.09        & 22.15         & -        & -                       \\
SDSS J213559.30+052700.5   &2015/10/06       & -6.97      & 16.92     & -      & -            & -        & -         & -        & -                       \\
V1239 Her                  &2013/04/13       & -64.09      & 39.55     & -      & -            & -        & -         & -        & -                       \\
                           &2017/04/25       & -138.80      & 39.63     & -54.69      & 34.59            & -27.22        & 31.59         & -        & -                       \\
                           &2017/05/23       & -124.70      & 39.61     & -49.10      & 37.20            & -        & -         & -        & -                       \\
NY Ser                     &2014/04/21       & -35.81      & 29.48     & -24.66      & 23.40            & -20.48        & 20.79         & -        & -                      \\
                           &2016/03/06       & -42.28      & 26.32     & -30.12      & 26.21            & -18.22        & 23.37         & -        & -                       \\
SDSS J153634.42+332851.9   &2015/05/12       & -185.50      & 33.03     & -93.91      & 34.49            & -30.12        & 24.47         &-         & -                        \\
GZ Cnc                     &2014/04/03       & -62.11      & 28.58     & -57.37      & 25.90            & -51.96        & 27.94         & -         & -                       \\
                           &2014/04/03       & -56.71      & 28.42     & -54.05      & 28.47            & -49.94        & 32.64         & -        & -                       \\
V344 Lyr                   &2016/05/07       & 4.66      & 33.89     & 7.29      & 23.97            & 7.70        & 21.69         & -        & -                       \\
SDSS J113950.58+455817.9   &2013/04/02       & -      & -     & -      & -            & -        & -         & -        & -                       \\
SDSS J155720.75+180720.2   &2017/03/27       & -17.40      & 30.80     & -35.91      & 36.00            & -22.67        & 34.00         & -        & -                       \\

\bottomrule
\end{tabular}
\end{table*}

\addtocounter{table}{-1}
\begin{table*}
\tiny
\centering
\caption{$-$continued.}
\label{Tab03}
\begin{tabular}{lccccccccc}
\toprule
\multirow{4}{*}{Target} & \multirow{4}{*}{Date-Obs}   & \multicolumn{2}{c}{H$\alpha$} & \multicolumn{2}{c}{H$\beta$} & \multicolumn{2}{c}{H$\gamma$}  & \multicolumn{2}{c}{HeII 4686} \\
\cmidrule(r){3-4} \cmidrule(r){5-6} \cmidrule(r){7-8}  \cmidrule(r){9-10}
&
&  $EW$      &  $FWHM$

&  $EW$      &  $FWHM$

&  $EW$      &  $FWHM$

&  $EW$      &  $FWHM$   \\
\midrule
AC LMi                     &2014/12/25       & -172.00      & 31.81     & -136.10      & 28.11            & -90.34        & 28.01         & -        & -                    \\
                           &2017/02/26       & -209.50      & 30.31     & -125.70      & 26.88            & -75.18        & 25.59         &  -       & -                       \\
V660 Her                   &2013/06/03       & -75.69      & 19.40     & -87.57      & 27.15            & -        & -         & -        & -                       \\
HY Psc                     &2015/10/08       & -206.10      & 13.98     & -158.60      & 14.04            & -        & -         & -        & -                       \\
WY Tri                     &2017/11/13       & -95.81      & 35.34     & -65.26      & 31.63            & -49.76        & 30.30         & -11.10        & 47.66                       \\
HT Cas                     &2017/12/17       & -170.40      & 45.30     & -78.52      & 41.14            & -43.43        & 37.14         &  -       & -                       \\
CC Cnc                     &2018/01/17       & -138.40      & 27.42     & -101.60      & 26.83            & -80.15        & 26.22         & -        & -                       \\
CY UMa                     &2011/12/19       & -123.90      & 29.67     & -66.63      & 25.69            & -37.01        & 23.02         & -        & -                       \\
                           &2013/04/28       & -74.11      & 35.53      & -      & -            & -        & -         & -        & -                       \\
                           &2018/01/12       & -154.40      & 28.67     & -93.26      & 24.89            & -63.28        & 23.02         & -        & -                       \\
SDSS J083845.23+491055.5   &2015/01/23       & -128.00      & 30.49     & -90.10      & 26.39            & -46.39        & 25.26         & -        & -                       \\
                           &2017/12/11       & -176.80      & 31.72     & -118.40      & 26.68            & -64.22        & 23.29         & -        & -                       \\
V1208 Tau                  &2014/10/06       & -56.56      & 29.69     & -43.99       & 30.67            & -41.25        & 22.71         & -        & -                       \\
V368 Peg                   &2015/10/03       & -39.63      & 24.99     & -22.80      & 19.22            & -        & -         & -        & -                       \\
IR Gem                     &2011/11/09       & -75.74      & 15.03     & -56.90      & 13.93            & -46.08        & 13.68         & -2.72        & 10.34                       \\
                           &2012/01/12       & -68.47      & 16.31     & -60.80      & 14.32            & -38.46        & 12.76         & -        & -                       \\
TY Psc                     &2011/11/10       & -      & -     & -      & -            & -        & -         & -        & -                                   \\
KS UMa                     &2015/02/18       & -89.54      & 22.12     & -65.77      & 19.85            & -41.41        & 18.06         & -        & -                       \\
SX LMi                     &2011/12/11       & -75.97      & 39.76     & -45.53      & 37.65            & -39.02        & 34.60         & -        & -                       \\
                           &2014/01/27       & -113.40      & 36.70     & -63.86      & 33.62            & -56.92        & 32.93         & -        & -                       \\
                           &2017/01/01       & -74.32      & 39.15     & -54.82      & 36.69            & -45.33        & 36.23         & -        & -                       \\
                           &2018/01/14       & -62.55      & 41.52     & -53.34      & 38.46            & -45.90        & 36.78         & -        & -                       \\
US 691                     &2016/01/09       & -76.51       & 42.54     & -39.69      & 39.08            & -        & -         & -        & -                       \\
                           &2016/12/22       & -75.14      & 43.69     & -56.31      & 42.60            & -        & -         & -        & -                       \\
                           &2016/12/26       & -55.64      & 38.05     & -43.42      & 32.44            & -25.98        & 29.64         & -        & -                       \\
GO Com                     &2013/04/13       & -223.30      & 31.29     & -114.10      & 27.61            & -62.93        & 27.20         & -        & -                       \\
                           &2017/04/25       & -128.50      & 32.78     & -95.55      & 29.47            & -71.00        & 27.55         & 11.31        & 44.75                       \\
2MASS J04260931+3541451    &2013/10/31       & -104.30      & 43.28     & -50.09      & 44.05            & -29.21        & 43.74         & -        & -                       \\
                           &2014/01/09       & -84.79      & 46.23     & -50.34      & 44.68            & -35.99        & 42.82         & -        & -                       \\
                           &2015/12/31       & -75.62      & 45.35     & -52.47      & 43.26            & -35.39        & 39.12         & -        & -                       \\
                           &2016/02/08       & -75.00      & 46.13     & -43.19      & 43.47            & -30.41        & 41.15         & -4.66        & 51.11                       \\
AK Cnc                     &2016/02/06       & -87.15      & 18.50     & -51.23      & 16.98            & -33.36        & 15.56         & -        & -                       \\
HW Boo                     &2018/01/17       & -140.50      & 25.20     & -116.40      & 22.80            & -73.67        & 20.99         & -        & -                       \\
ER UMa                     &2016/04/12       & -48.87      & 21.93     & -42.26      & 23.74            & -36.36        & 24.10         & -6.91        & 37.23                        \\
                           &2016/12/01       & -20.65      & 19.04     & -8.13      & 14.30            & -4.69        & 13.13         & -        & -                       \\
MR UMa                     &2013/05/14       & -89.73      & 22.97     & -40.45      & 19.49            & -8.43        & 12.45         & -        & -                       \\
DW Cnc                     &2011/12/19       & -135.40      & 27.15     & -126.20      & 23.17            & -98.99        & 21.65         & -        & -                       \\
                           &2014/11/04       & -166.20      & 27.27     & -140.40      & 22.85            & -108.60        & 21.25         & -        & -                        \\
                           &2015/12/21       & -199.90      & 27.30     & -147.90      & 22.05            & -109.70        & 20.06         & -        & -                       \\
                           &2015/12/21       & -189.50      & 26.34     & -154.30      & 21.39            & -117.80        & 19.72         & -        & -                       \\
FS Aur                     &2011/11/10       & -31.19      & 20.54     & -30.00      & 15.60            & -26.05        & 13.92         & -        & -                       \\
                           &2013/03/06       & -30.92      & 20.94     & -45.43      & 19.05            & -35.65        & 16.76         & -        & -                       \\
EZ Lyn                     &2013/01/08       & -56.97      & 43.94     & -13.37      & 30.64            & -        & -         & -        & -                       \\
NZ Boo                     &2014/05/22       & -190.60      & 47.80     & -73.58      & 41.62            & -55.03        & 41.24         & -        & -                       \\
EG Cnc                     &2013/01/15       & -111.00      & 53.52     & -10.66      & 35.09            & -        & -         & -        & -                            \\
                           &2014/12/24       & -118.30      & 51.08     & -      & -            & -        & -         & -        & -                       \\
                           &2016/11/24       & -130.00      & 55.28     & -      & -            & -        & -         & -        & -                          \\
RZ LMi                     &2013/11/12       & -      & -     & 6.88      & 41.84            & 6.13        & 32.78         & -        & -                       \\
V355 UMa                   &2013/04/11       & -92.49      & 30.49     & -18.51      & 21.99           & -        & -         & -        & -                       \\
                           &2017/04/27       & -102.80      & 30.59     & -29.61      & 23.34            & -11.23        & 19.40         & -        & -                       \\
PQ And                     &2017/09/22       & -75.46      & 30.38     & -16.80      & 19.33            & -        & -         & -        & -                       \\
V627 Peg             &2016/10/03       & -27.10      & 28.28     & -14.92      & 22.79            & -17.05        & 23.43         & -        & -                       \\
EI Psc                     &2013/09/25       & -20.34      & 37.68     & -12.32      & 26.85            & -19.13        & 36.68         & -        & -                       \\
YZ LMi                     &2016/01/07       & -13.30      & 32.69     & -      & -            & -        & -         & -        & -                       \\
                           &2016/12/24       & -      & -     & -      & -            & -        & -         & -        & -                       \\
PTF1 J071912.13+485834.0   &2016/12/19       & -      & -     & -      & -            & -        & -         & -        & -                      \\
UCAC3 232-65219            &2016/01/18       & -24.70      & 26.38     & -16.87      & 26.06            & -19.79        & 26.64         & -        & -                       \\
SDSS J003303.94+380105.4   &2015/10/16       & -70.21      & 21.26     & -62.38      & 26.99            & -32.33        & 28.00         & -        & -                       \\
                           &2015/10/17       & -62.12      & 24.38     & -41.33      & 28.21            & -40.00        & 29.89         & -        & -                       \\
SDSS J084358.08+425036.8   &2016/02/08       & -      & -     & -      & -            & -        & -         & -        & -                            \\
SDSS J090628.24+052656.9   &2016/03/05       & -4.23      & 18.41     & -      & -            & -        & -         & -        & -                        \\
SDSS J124417.89+300401.0   &2017/02/01       & -172.10      & 20.65     & -30.32      & 15.53           & -        & -         & -        & -                        \\
SDSS J153015.04+094946.3   &2017/04/25       & -60.84      & 31.59     & -43.13      & 26.86            & -        & -         & -        & -                         \\
\bottomrule
\end{tabular}
\end{table*}

\subsection{Dwarf novae in outburst}

The spectra in Figs. 8-11 shows prominent characteristics of an optically thick disc, typical of dwarf novae in outburst. These outburst spectra reveal a variety of characteristics from object to object, and the same object in different stages of outburst is also widely different. In general, the spectrum in outburst displays a bluer continuum than in quiescence, and the absorption lines will become dominant. Here we gave some notes and comments on the outburst spectra below.

\emph{AR And}. The spectrum taken on 2016 Jan 5 shows broad Balmer absorption lines, as well as a weak H$\alpha$ emission line. The AAVSO light curve appears to show the rise into an outburst peak. This spectrum is similar to U Gem at outburst peak (see Fig. 1) but without HeII $\lambda$4686 line. Shafter \& Veal (1995) published several outburst spectra. However, their spectral coverage is quite limited, only $~$800 {\AA} (i.e. 4300-5000{\AA}).

\emph{AT Cnc}. The spectrum shows an unusual appearance, which is unlike its outburst spectrum published by Morales-Rueda \& Marsh (2002). The spectrum was taken on 2013 April 14, corresponding to the outburst peak in the AAVSO light curve. Clearly, the peculiar shape of the continuum is not real.

\emph{AY Psc}. This object was observed on three different nights. The spectrum on 2015 Oct 15 is at outburst peak confirmed by examining AAVSO light curve, and exhibits broad Balmer emission lines and several weak neutral helium emission lines (e.g. HeI $\lambda$4914, 5876, 6678). We note that the strong HeII $\lambda$4686 and Bowen blend lines (CIII/NIII $\lambda$4634-4651) were detected in emission, indicating the system with quite high mass transfer rate or a spiral structure in the accretion disc. There are also some absorption features at HeI $\lambda$4026 and 4471, FeI $\lambda$5018, FeII $\lambda$4924, Mg $\lambda$5178 and Na D lines. The activity state of the spectrum obtained on 2015 Dec 12 is identified as the early decline. This spectrum is similar to the outburst spectrum mentioned above, but the strength of the absorption lines decreased slightly. Third spectrum was observed on 2017 Dec 14, it shows an appearance of a quiescent continuum and H$\alpha$, H$\beta$ and HeII emission lines. This observed date correspond to a short standstill in the AAVSO light curves. However, the shape of the spectrum is completely different from outburst spectra, and its flux value is very peculiar. Therefore, it is not real and most likely attributable to a bad calibration. Besides, the quiescent spectra (the wavelength coverage 4250-4950 {\AA}) of AY Psc were previously presented by Szkody \& Howell(1993), but the outburst spectra have not yet been found in the literature and the SDSS database. The LAMOST provides the first outburst spectral data of this object.

\emph{ER UMa}. The spectrum obtained on 2016 Dec 1 exhibits a blue continuum and strong Balmer emission lines but lacks absorption lines. There are also weak HeI and HeII $\lambda$4686 lines in emission. By examining AAVSO data, however, we find that the spectrum is at outburst peak. This object is a peculiar SU UMa-type dwarf nova which is called as the ER UMa-type star, displaying much shorter supercycles (Kato \& Kunjaya 1995; Robertson et al. 1995). The supercycles were explained as the eccentric dissipation of accretion disc (Hellier 2001). It is still uncertain whether these strong Balmer emission lines are related to the eccentric disc. Morales-Rueda \& Marsh(2002) previously published an outburst spectrum with weak HeII $\lambda$4686 line. While the Balmer lines were observed in absorption with core emission in this spectrum, their emission lines such as observed in the LAMOST outburst spectrum were not seen due to the strong continuum during outburst.

\emph{IU Leo}. This system was observed on 2015 Jan 1 and displays typical spectrum in outburst. The outburst stage is determined as the early decline. The spectrum shows a steep continuum in the blue and the broad absorption wings around weak Balmer emission lines, implying that it is dominated by the optically thick accretion disc. In addition, some weak metal absorption lines such as Mg $\lambda$5176 and Na D lines are also recognized. Its quiescent spectra were reported by Thorstensen et al. (2010), but the outburst spectra did not appeared in the literature so far.

\emph{RZ LMi}. This object is an ER UMa-type star, and also shows short supercycles in the AAVSO light curve. The spectrum taken on 2013 Nov 12 at V=15.0 mag, shows broad Balmer absorption lines with the emission core, typical of a dwarf nova in outburst. The activity state was confirmed as the decline from a superoutburst. Szkody \& Howell(1992) reported a 4200-7800 {\AA} spectrum while RZ LMi was at V=14.7 mag. Compared with the LAMOST spectrum, it was closer to the outburst peak and only H$\alpha$ line has an emission core.

\emph{TW Tri}. The spectra were obtained near outburst peak (2013 Sep 25) and on the rise (2015 Oct 1), respectively. The former exhibits typical spectrum of a dwarf nova in outburst, similar to other two U Gem-type systems AR And and U Gem. There are also HeII $\lambda$4686 emission with Bowen blend lines, and weak Mg $\lambda$5176 and Na D absorption lines. The later shows a rising blue continuum with the absorption wings around Balmer emissions. The spectroscopic studies given by Zwitter \& Munari (1994) and Thorstensen et al.(1998) reported the quiescent spectra of this object, but the outburst spectra have not been found in the literature as yet.

\emph{X Leo}. The spectrum observed on 2016 Nov 27 was identified as the decline from an normal outburst (V=13 mag). It has a steep blue continuum and shows an absorption feature around the Balmer emission lines, which are most likely due to the optically thick disc. The outburst spectra, covering $~$4200-5000 {\AA}, have appeared in Szkody et al. (1990) and Morales-Rueda \& Marsh (2002). The former displayed H$\beta$, H$\gamma$ absorption lines with a very weak emission core, and was obtained near outburst peak. The latter was observed at V=13.1 mag and revealed an absorption feature around the H$\beta$, H$\gamma$ lines, which are similar to our spectrum in the range of 4200-5000 {\AA}.

\emph{V344 Lyr}. This system was detected near outburst maximum, revealing a bluer continuum with broad Balmer and helium absorption lines. Although the AAVSO light curves are poor, the spectrum is without a doubt near the outburst peak. Previous spectrum at quiescent was presented in Howell et al.(2013) but the spectra at outburst was not reported in the literature.

\emph{PTF1 J071912.13+485834.0}. This star is an AM CVn system, and the spectra at quiescent and outburst were presented in Levitan et al. (2011). Our spectrum obtained on 2016 Dec 19 shows a rising blue continuum but other features are not apparent due to a poor signal-to-noise ratio. 

\emph{SDSS J081256.85+191157.8}. The spectrum shows typical signature of an optically thick disc, indicating a CV in outburst or a NL with high accretion rate. This object was discovered in SDSS data (Szkody et al. 2006), and the SDSS spectrum has a similar appearance with the LAMOST spectrum. Further spectroscopic study from Thorstensen et al.(2015) suggested it may be a NL. Due to the lack of AAVSO data, the activity state of the system cannot be determined at present. The prominent HeII $\lambda$4686 with Bowen blend lines and the CRTS light curve reveal it to be a NL as expected.

\emph{SDSS J090628.24+052656.9}. The spectrum taken on 2016 March 5 shows the blue continuum plus weak H$\alpha$ and the absorption wings around H$\beta$ emmision, typical of an dwarf nova during rising or decline. The SDSS spectrum at outburst was presented by Pourbaix et al.(2005) and displayed the blue continuum and broadened Balmer absorption lines, suggesting that it is at outburst peak. 
Some unidentified emission lines around $\lambda$6893 and $\lambda$7674 were also seen in its LAMOST spectrum. These features are near to major telluric features and we have not found any sign of these emission lines in its non-telluric-corrected spectrum. Thus, they most likely attributable to a telluric-corrected issue.

\emph{SDSS J213559.30+052700.5}. The spectrum obtained on 2015 Oct 6 displays a steep continuum with broad absorption features around weak Balmer emission core, typical of dwarf novae in outburst or NLs in a high state. The light curve from CRTS revealed that this star is a VY Scl-type NL (Drake et al. 2009). 
A spectrum in its low state was presented by Carter et al.(2013), but the spectra in high state was not found in the literature so far.

\emph{V537 Peg}. The spectrum on 2014 Nov 11 is obtained for the first time, showing a sharp blue continuum plus broad absorption features around strong Balmer lines. There are some relatively weak emission such as HeI series, HeII $\lambda$4686 and the calcium triplet lines. The Na D absorption lines are visible in the spectrum. Compared with the spectra of U Gem-type stars noted above, this spectrum has stronger Balmer lines. Of course, this system also has a longer orbital period (10.14 hours) than others (Bernhard et al. 2007).

\emph{SY Cnc}. Some spectroscopic investigations of SY Cnc were performed and its spectral features were reported in the literature (e.g. Herbig 1950; Oke \& Wade 1982; Williams 1983; Morales-Rueda \& Marsh 2002; Casares et al.2009). The LAMOST spectra of SY Cnc are observed on seven different nights (see Fig. 11), and display typical outburst signature. In addition to the spectrum taken on 2018 Dec 3, others were obtained from 2017 Dec 18 to 2018 Jan 12. This offers a good opportunity to trace its spectral evolution at different outburst stages. Large EW variations at the different stages were detected in the Balmer lines (see Table 2). This object is a Z Cam star with a long orbital period of $\sim$ 9.18 hours, and exhibits sinusoidal outbursts (Casares et al. 2009). The AAVSO light curve shows that the spectra on 2017 Dec 18 and 21 are at late decline (lower panel of Fig. 11). As the outburst fades, the continuum becomes flatter and the strength of Balmer lines increases obviously. We also note that the broad absorption wings nearly disappear. The spectra obtained from 2018 Jan 8 to 12 were confirmed to be at early decline of another outburst. The main change is still stronger Balmer emission feature with the decline, revealing a cooler accretion disc. The last spectrum on 2018 Dec 3 shows broad absorption wings around weak Balmer emission. Although this date does not covered by the AAVSO data, its state can also be assessed as outburst peak. Other interesting features in these spectra are clear HeII $\lambda$4686 with Bowen blend emission lines and the Na D lines in absorption.

\begin{figure}
\begin{center}
\includegraphics[width=1.0\columnwidth]{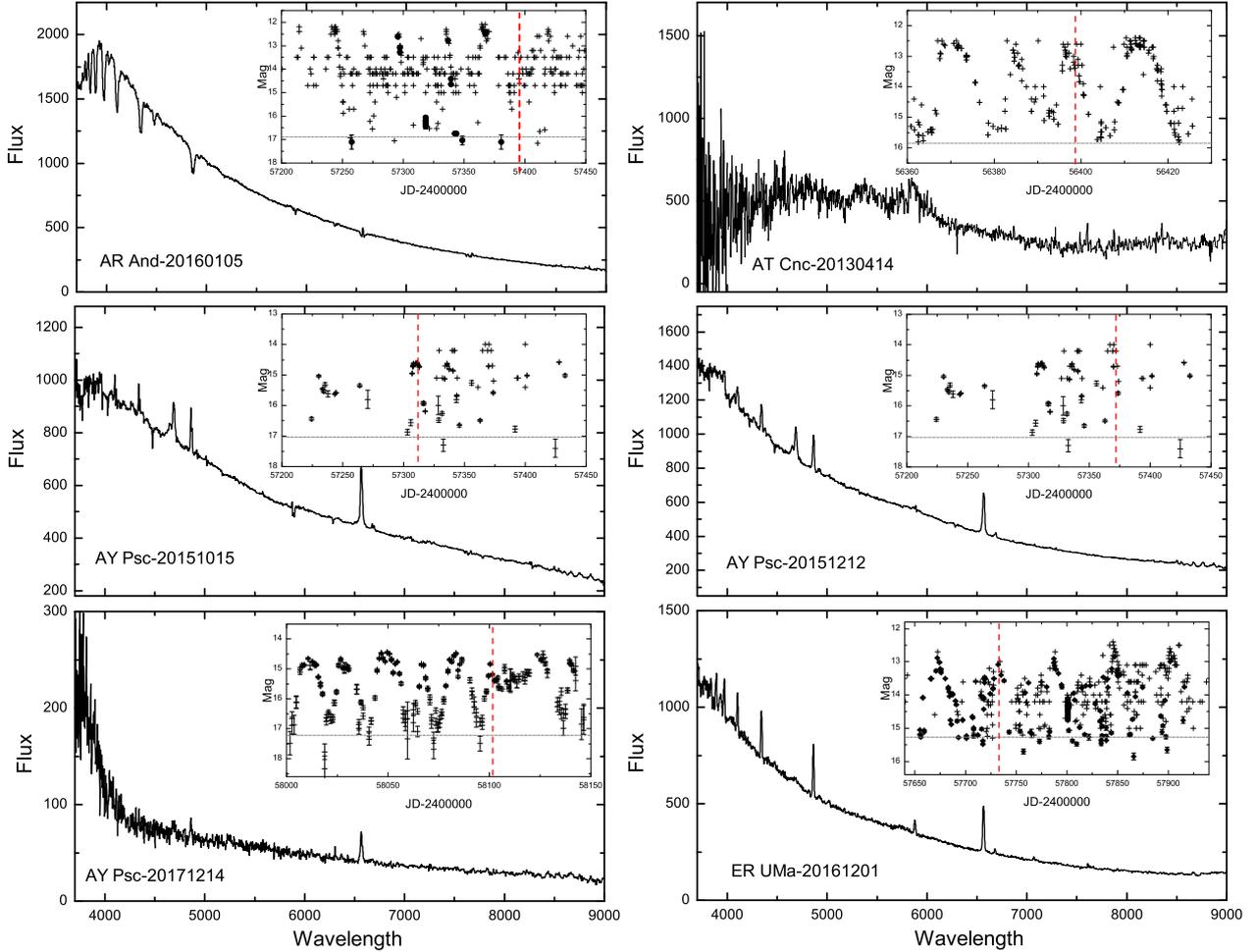}
\caption{LAMOST outburst spectra of four dwarf novae: AR And, AT Cnc, AY Psc and ER UMa. The AAVSO light curves corresponding to the spectra are plotted in the inset, and a red dashed line denotes the dates for observing the spectrum.}
\end{center}
\end{figure}

\begin{figure}
\begin{center}
\includegraphics[width=1.0\columnwidth]{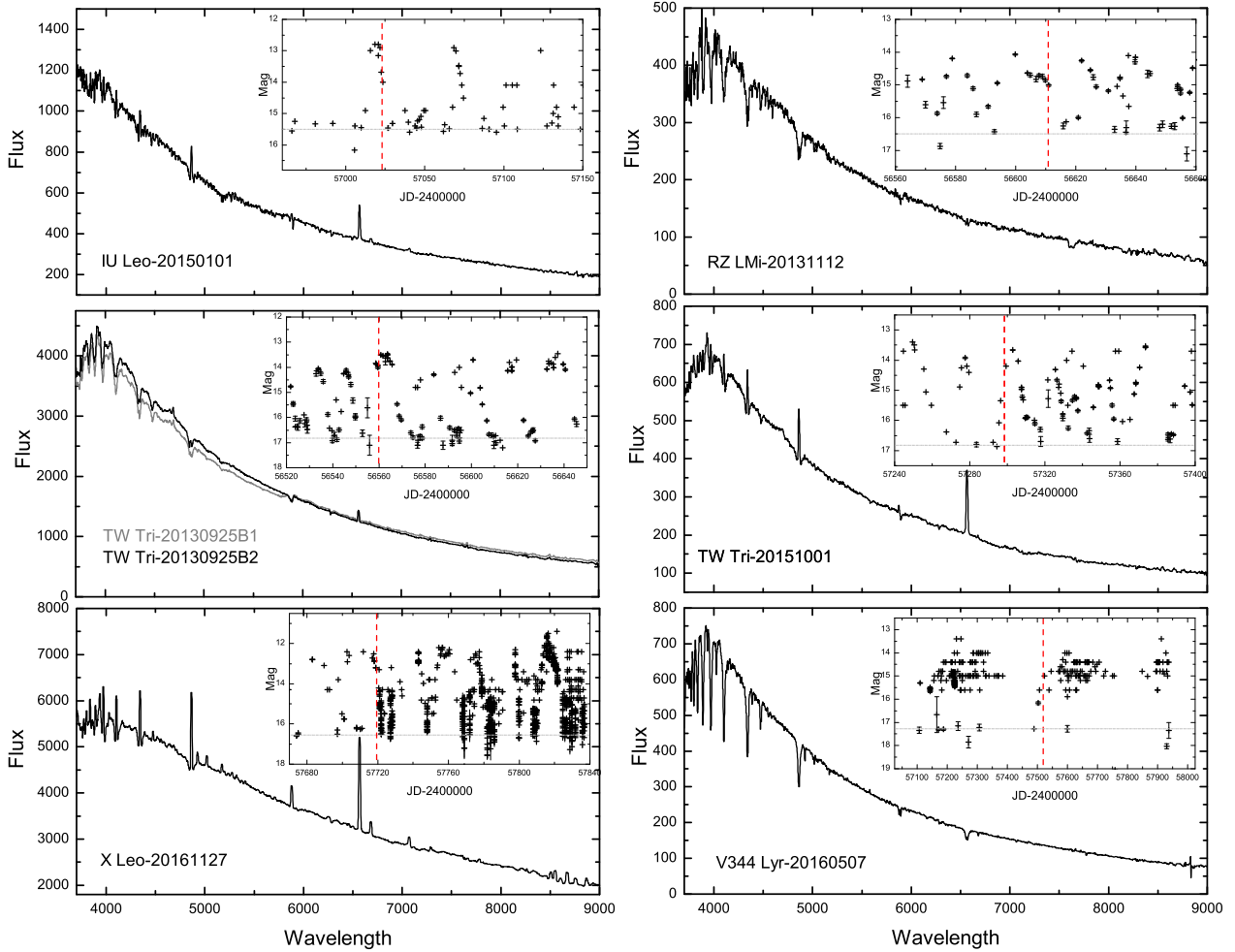}
\caption{LAMOST outburst spectra of five dwarf novae: IU Leo, RZ LMi, TW Tri, X Leo and V344 Lyr. Other details are the same as the caption of Fig. 1.
 }
\end{center}
\end{figure}

\begin{figure}
\begin{center}
\includegraphics[width=1.0\columnwidth]{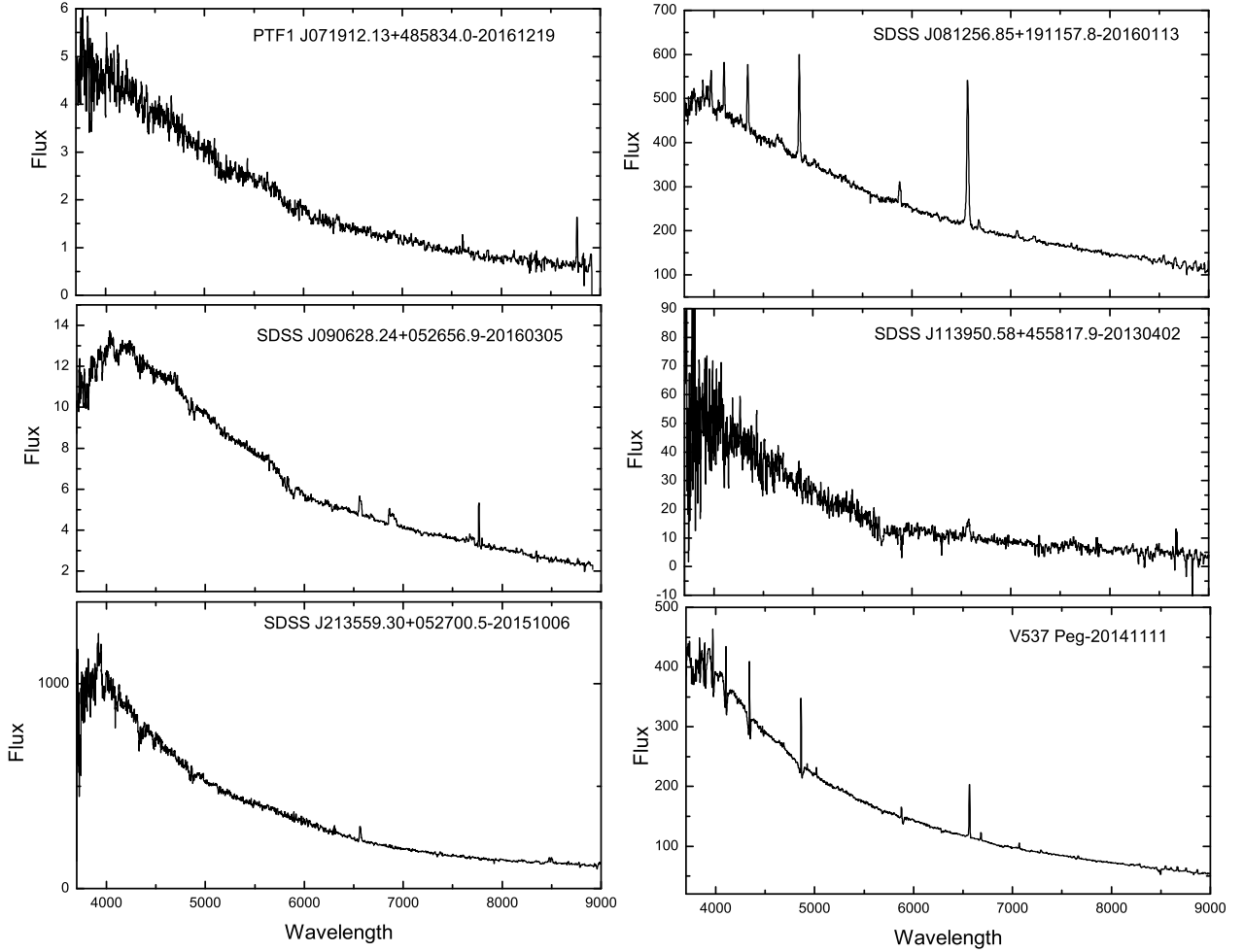}
\caption{LAMOST spectra of six objects: PTF1 J071912.13+485834.0, SDSS J081256.85+191157.8, SDSS J090628.24+052656.9, SDSS J113950.58+455817.9, SDSS J213559.30+052700.5 and V537 Peg, typical of dwarf novae in outburst or NLs at the high state. These spectra don't covered by the AAVSO data.}
\end{center}
\end{figure}

\begin{figure}
\begin{center}
\includegraphics[width=1.0\columnwidth]{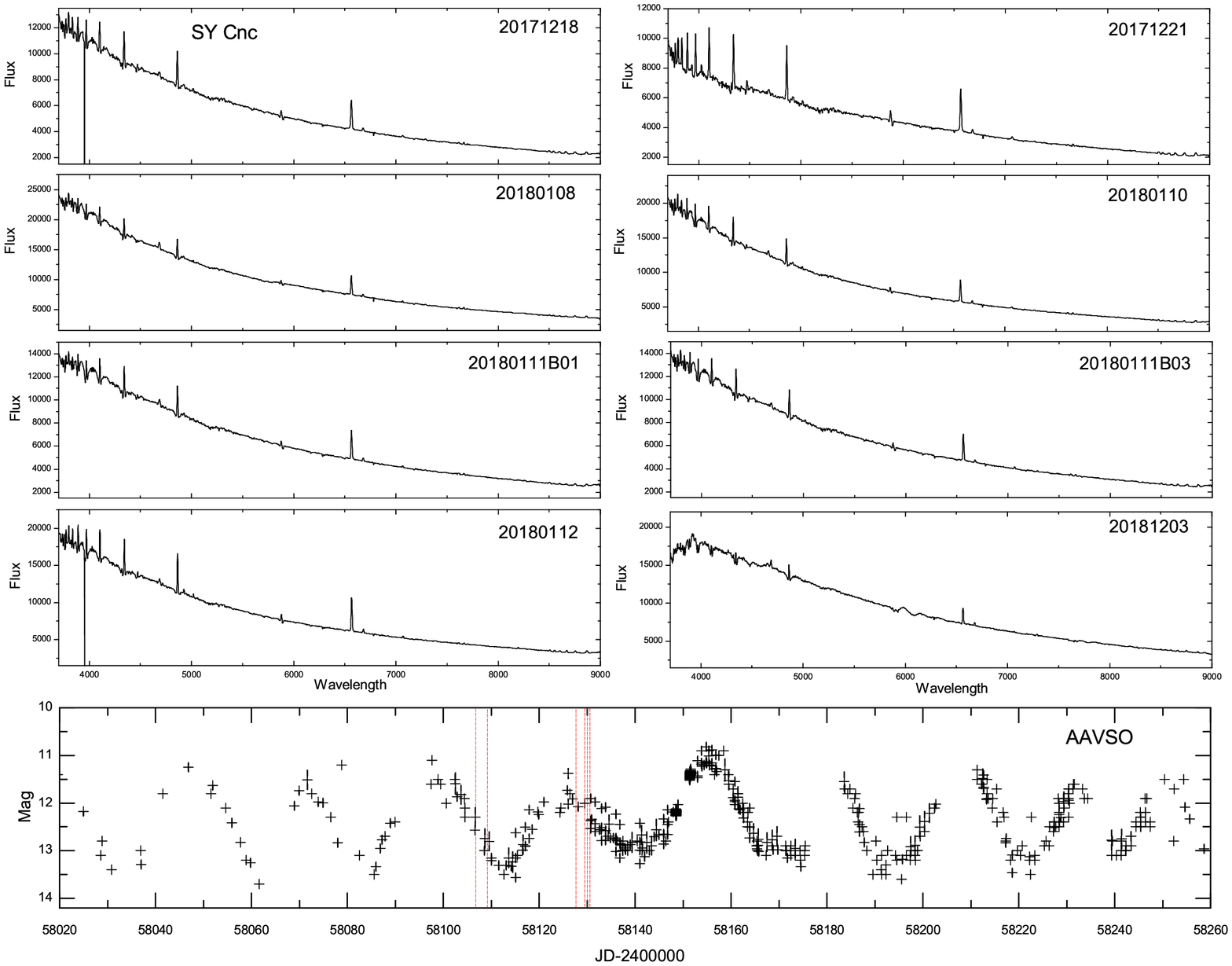}
\caption{LAMOST outburst spectra of SY Cnc obtained on seven different nights. The AAVSO light curve was plotted in the bottom panel, indicating the quasi-periodic outburst. Six red dashed lines indicate the dates that the spectra were observed on the first six nights. The spectrum taken on 2018 Dec 3 dose not covered by the AAVSO data.
 }
\end{center}
\end{figure}

\section{Discussion and Conclusion}

Over the years, the number of dwarf nova-type CVs is growing rapidly due to modern astronomical surveys. Most of information regarding dwarf novae mainly focus on their photometric variability. However, to study the intrinsic properties of the binary components in more detail, the optical spectroscopy is a more effective means. Of all 681 known dwarf novae listed in RKcat 7.24 and collected from the literatures, 76 were detected by LAMOST and a total of 131 spectra were obtained. Among the 76 objects, about 20 display typical spectra of dwarf novae in or near outburst and only 2 are the first spectra of this object. By checking the spectra of same objects from the SDSS data, we find that 36 have been observed by SDSS. In this paper, we presented the LAMOST spectra of 76 dwarf novae and studied their spectral properties in quiescence and outburst.

In most dwarf novae, the accretion disc emission will contribute the most flux in the total spectrum.
The quiescent spectra shown in Figs. 2-7 are dominated by strong Balmer and neutral helium series in emission, which superposed on a relatively flat continuum. The main reason is that in quiescence the dwarf novae have an optically thin disc with a large temperature range from the boundary layer to outer edge. However, in low accretion rate systems the discs are faint, and the signatures of white dwarf and donor can be found in the spectra. 10 of our objects display the M-type donor features, which usually characterise strong atomic or molecular absorption bands (e.g., Mg and TiO). The spectral features of K-type donor are also visible in two famous dwarf novae, GK Per and SS Cyg. Additionally, four WZ Sge-type stars with an extremely low mass transfer rate have typical white-dwarf-dominated spectra, which often reveal a steep blue continuum plus broad absorption wings around the Balmer emissions. It was widely believed that the absorption features result from the white dwarf's atmosphere. Apart from these, the Na D absorption lines ($\lambda$5890/5896{\AA}) can be found in some spectra. We suggest that the Na D lines root in a combination of interstellar absorption and/or the donor star. In particular, the spectrum of V367 Peg was observed for the first time, and shows evidence of a M-type donor. The donor's spectral type was estimated as M3-5 based on the spectral features and its orbital period. Other particular systems and features also have been noted and discussed in the section 3.1. Further follow-up observations will be needed to study these unusual spectra in the future.

In contrast to the quiescent state, the accretion disc in outburst turns into optically thick and the absorption lines become the dominant features. Moreover, the continuum in outburst is bluer than in quiescence due to the hotter disc. The outburst spectra are shown in Figs. 8-11 and the majority of these were covered by the AAVSO data. In these spectra, the different targets show a variety of spectral features and the same object at different outburst stages also varied markedly. Among those objects, 6 have the first outburst spectra. We also compared other outburst spectra with the published spectra. The spectra of three U Gem-type stars, AR And, TW Tri and U Gem, at the outburst peak reveal strong absorption lines, expect for a weak H$\alpha$ line in emission. Several long-period systems near the outburst peak display prominent disc emission features such as in AY Psc, V537 Peg and SY Cnc. All these indicate that the absorption features are not an unique indicator of the outburst spectra, and there may be some emitting regions in the optically thick disc. The strength of emission lines could be closely connected to the mass transfer rate. The outburst spectra of two ER UMa-type stars were also reported. All outburst data presented here will be useful for the spectroscopic studies of dwarf novae due to wide wavelength coverage. It is worth mentioning that the spectra of seven objects in outburst show clear HeII $\lambda$4686 emission line along with the Bowen blend. This feature is a good tracer of spiral arms of accretion disc in dwarf novae (Harlaftis et al. 1999; Smak 2001; Ogilvie 2002). Among these seven objects with HeII $\lambda$4686 emission, the spiral structure in U Gem has been detected in previous study (see Groot 2001). Other six systems: AY Psc, ER UMa, SDSS J081256.85+191157.8, SY Cnc, TW Tri and V537 Peg, are good candidates for probing the spiral asymmetries. However, it is important to point out that there are additional challenges to explore the spiral structure by HeII $\lambda$4686 emission, because the outbursts are a transient process, it is required to notice the outburst state and then to obtain the spectra rapidly.

\bigskip

\vskip 0.3in \noindent
This work is partly supported by the National Natural Science Foundation of China (Nos. 11803083, U1831120, 11573063, 11611530685, U1731238), the Key Science Foundation of Yunnan Province (No. 2017FA001) and the West Light Foundation of The Chinese Academy of Sciences (No. E0290601). The spectral data presented in this paper were taken from the LAMOST database. The LAMOST survey is funded by the National Development and Reform Commission. The telescope is operated for the National Astronomical Observations, Chinese Academy of Sciences. We also appreciate the contributions of the Ritter-Kolb catalogue (RKcat 7.24), the AAVSO database and the International Variable Star Index database (VSX\footnote{https://www.aavso.org/vsx/index.php?view=search.top}) in this research.


\end{document}